\documentclass[format=acmsmall, review=false, screen=true]{acmart}
\usepackage{booktabs} % For formal tables
\usepackage{makecell}
\usepackage{lipsum}
\usepackage[utf8]{inputenc}
\usepackage{array}
\usepackage{wrapfig}
\usepackage{multirow}
\usepackage{tabularx}
\usepackage{pifont}
\usepackage{lineno}
\usepackage{balance}
\usepackage{xcolor,pifont}
\usepackage{ragged2e}
\usepackage{hyphenat} % 添加断字支持
\usepackage{pifont} % 提供 \ding 命令

\usepackage{soul}

\usepackage[linesnumbered, ruled,vlined]{algorithm2e}
\RequirePackage{pdflscape}
\usepackage{flushend}
\usepackage{multicol}
\usepackage{placeins}
\usepackage{booktabs}
\usepackage{indentfirst}
\usepackage{fancybox}
\usepackage[most]{tcolorbox}
\usepackage{fontawesome}
\usepackage{mathtools}
\usepackage{MnSymbol,wasysym}
\usepackage{rotating}
\usepackage{adjustbox}
\usepackage{colortbl}
\usepackage{color}
\usepackage{tcolorbox}
\usepackage{enumitem}
\newtcolorbox{mybox}[2][]
{colback = white, colframe = black, fonttitle = \bfseries,
    colbacktitle = gray, enhanced,
    attach boxed title to top left={yshift=-3mm, xshift=3mm},
    title=#2, #1}

\usepackage{framed}
\definecolor{Gray}{gray}{0.9}
\definecolor{shadecolor}{gray}{0.95}
\usepackage{tikz}
\usetikzlibrary{trees,positioning,shapes,shadows,arrows}
\tikzset{
  basic/.style  = {draw, text width=2cm, drop shadow, font=\sffamily, rectangle},
  root/.style   = {basic, rounded corners=2pt, thin, align=center, fill=white},
  level-2/.style = {basic, rounded corners=6pt, thin,align=center, fill=white, text width=3cm},
  level-3/.style = {basic, thin, align=center, fill=white, text width=1.8cm}
}
\newcommand{\todo}[1]{}
\renewcommand{\todo}[1]{{\color{red} TODO: {#1}}}
\usepackage[labelformat=simple]{subcaption} % 简化标签格式
\newcommand{\ours}[1]{\textsc{MAAD}}

\usepackage[normalem]{ulem}

\begin{document}

\title{MAAD: Automate Software Architecture Design through Knowledge-Driven Multi-Agent Collaboration}

\author{Ruiyin Li}
% \authornotemark[1]
\email{ryli_cs@whu.edu.cn}
\affiliation{%
  \institution{School of Computer Science, Wuhan University}
  \city{Wuhan}
  \country{China}
}

\author{Yiran Zhang}
\email{yiran002@e.ntu.edu.sg}
\affiliation{
  \institution{Nanyang Technological University}
  \city{Singapore}
  \country{Singapore}
}

\author{Xiyu Zhou}
\email{xiyuzhou@whu.edu.cn}
\affiliation{
  \institution{School of Computer Science, Wuhan University}
  \city{Wuhan}
  \country{China}
}

\author{Peng Liang}
\email{liangp@whu.edu.cn}
%\authornote{Corresponding author}
\affiliation{
  \institution{School of Computer Science, Wuhan University}
  \city{Wuhan}
  \country{China}
}

\author{Weisong Sun}
\email{weisong.sun@ntu.edu.sg}
\affiliation{
  \institution{Nanyang Technological University}
  \city{Singapore}
  \country{Singapore}
}

\author{Jifeng Xuan}
\email{jxuan@whu.edu.cn}
\affiliation{
  \institution{School of Computer Science, Wuhan University}
  \city{Wuhan}
  \country{China}
}

\author{Zhi Jin}
\email{zhijin@pku.edu.cn}
% \authornote{Corresponding author}
%\authornotemark[1]
\affiliation{
  \institution{School of Computer Science, Wuhan University}
  \city{Wuhan}
  \country{China}
}

\author{Yang Liu}
\email{yangliu@ntu.edu.sg}
\affiliation{
  \institution{Nanyang Technological University}
  \city{Singapore}
  \country{Singapore}
}

\thanks{This research is supported by the National Natural Science Foundation of China (NSFC) with Grant No. 62402348 and 62172311; National Research Foundation, Prime Minister's Office, Singapore under the Campus for Research Excellence and Technological Enterprise (CREATE) Programme; the National Research Foundation, Singapore, and DSO National Laboratories under the AI Singapore Programme (AISG Award No: AISG2-GC-2023-008). The authors would also like to thank the architects who participated in the interviews in this study.}

\acmJournal{TOSEM}
\acmVolume{0}
\acmNumber{0}
\acmArticle{0}
\acmMonth{0}

\renewcommand{\shortauthors}{Li et al.}

\begin{abstract}
Software architecture design is a critical, yet inherently complex and knowledge-intensive phase of software development. It requires deep domain expertise, development experience, architectural knowledge, careful trade-offs among competing quality attributes, and the ability to adapt to evolving requirements. Traditionally, this process is time-consuming and labor-intensive, and relies heavily on architects, often resulting in limited design alternatives, especially under the pressures of agile development. 
While Large Language Model (LLM)-based agents have shown promising performance across various software engineering tasks, their application to architecture design remains relatively scarce and requires more exploration, particularly in light of diverse domain knowledge and complex decision-making. In addition, relying on a single LLM agent often leads to unreliable and inconsistent outcomes, especially when handling tasks that require collaborative reasoning and deliberation. 
To address the challenges, we proposed MAAD (Multi-Agent Architecture Design), an automated framework that employs a knowledge-driven Multi-Agent System (MAS) for architecture design. 
MAAD orchestrates four specialized agents (i.e., \textit{Analyst}, \textit{Modeler}, \textit{Designer} and \textit{Evaluator}) to collaboratively interpret requirements specifications and produce architectural blueprints enriched with quality attributes-based evaluation reports. 
We then evaluated MAAD through a case study and comparative experiments against MetaGPT, a state-of-the-art MAS baseline. Our results show that MAAD's superiority lies in generating comprehensive architectural components and delivering insightful and structured architecture evaluation reports. Feedback from industrial architects across 11 requirements specifications further reinforces MAAD's practical usability. We finally explored the performance of the MAAD framework with three LLMs (GPT-4o, DeepSeek-R1, and Llama 3.3) and found that GPT-4o exhibits better performance in producing architecture design, emphasizing the importance of LLM selection in MAS-driven architecture design.
\end{abstract}
% [---- For Submission Only -----]
% Software architecture design is a complex, knowledge-intensive phase requiring domain expertise, experience, and careful trade-offs among quality attributes, while adapting to changing requirements. Traditional approaches are time-consuming, depend heavily on architects, and often yield limited design alternatives—particularly under agile development pressures. Although Large Language Model (LLM) agents show promise in software engineering tasks, their use in architecture design remains underexplored, especially given the need for diverse knowledge and collaborative decision-making. Single-agent solutions frequently produce inconsistent results. We propose MAAD (Multi-Agent Architecture Design), a knowledge-driven multi-agent framework that coordinates four specialized agents (i.e., Analyst, Modeler, Designer, and Evaluator) to interpret requirements and generate architectural artifacts accompanied by quality-based evaluation reports. A case study and comparative experiments against MetaGPT, a leading MAS baseline, demonstrate that MAAD delivers more comprehensive architecture components and structured evaluations. Feedback from industrial architects reviewing requirements specifications generated by MAAD further confirms its practical applicability. Additionally, we evaluated MAAD using three LLMs (i.e., GPT-4o, DeepSeek-R1, and Llama 3.3) finding that GPT-4o performs best in producing effective designs, highlighting the impact of LLM selection on MAS-driven architecture design.

 % [CCS的分类] https://dl.acm.org/ccs

\maketitle

\section{Introduction}\label{sec:Introduction}

Software architecture design lies at the heart of any successful software project. It defines the system's high‑level structure, allocates responsibilities to components, and prescribes the interactions that satisfy both functional requirements and quality attributes~\cite{Bass2021SAP}. In practice, architects must translate often‐ambiguous requirements into concrete modules and connectors, select appropriate architectural patterns, and balance competing concerns on quality attributes (e.g., performance, security, maintainability)~\cite{Garlan2009ArchMism}. This process is inherently knowledge‑intensive and demands deep domain expertise, extensive engineering experience, and careful trade‑off analysis. As requirements evolve or new constraints emerge (e.g., legacy dependencies, regulatory mandates, or shifting business goals), the architecture must adapt to the latest requirements without undermining system integrity. Such complexity often leads to cognitive overload, reliance on tacit personal knowledge, and limited exploration of alternative designs, creating bottlenecks that delay delivery and hinder the ability to scale design efforts across projects.

The advent of Large Language Models (LLMs) has profoundly revolutionized the landscape of Software Engineering (SE) practices, introducing a new paradigm that integrates Generative AI (GenAI) into various development workflows. Especially, LLM-based tools such as ChatGPT are already being adopted across a broad spectrum of SE activities \cite{li2025ChatGPT, ChatGPT2025SLR}. However, when a single LLM agent tackles complex, multi‐step tasks, the results can be unreliable or inconsistent, as isolated reasoning tends to overlook cross‐cutting concerns and introduce \textit{hallucinations} \cite{Zhang2025Hallucination}. To mitigate these issues, recent research has turned to Multi‑Agent Systems (MAS), where several role‑specific LLM agents communicate and refine outputs. By emulating the collaborative dynamics of human design teams, each agent focuses on a particular subtask, such as requirements analysis~\cite{jin2024mare, He2025mas}. 

Unlike single LLM deployments, MAS architectures emulate human development teams by enabling agents to focus on specific roles, reason independently, and communicate iteratively toward a shared goal~\cite{He2025mas}. This distributed intelligence approach excels in scenarios that require diverse expertise, complex problem decomposition, parallel task execution, and collaborative deliberation. As shown in recent work~\cite{sun2025tdlh, du2024improving, talebirad2023multi}, MAS-based GenAI frameworks can reduce hallucinations, increase robustness, and improve overall productivity, especially in dynamic environments where rapid iteration is essential. These capabilities are especially critical in the context of modern software development, where the accelerated pace of software delivery demands rapid iteration and market responsiveness to evolving market requirements.

However, despite promising advances, the high-abstraction and knowledge-intensive nature of software architecture design remains under‑automated. Existing LLM and MAS applications excel at coding tasks but receive comparatively less focus on the architectural phase, where decisions about module decomposition, protocol selection, and non‑functional trade‑offs are both interdependent and domain‑specific. Architects still carry the bulk of this work, leading to several challenges: limited reuse of prior designs when diving into unfamiliar domains, low generalizability of the architecting process due to the difficulty of applying and transferring knowledge across teams, and insufficient integration of domain-specific external knowledge. As a result, organizations struggle to accelerate the architecture formulation and maintain consistency as projects evolve over time and grow in scope.

To bridge this gap, our \textbf{\textit{goal}} is to propose and validate an automated software architecture design framework \textbf{Multi-Agent Architecture Design (MAAD)}, which orchestrates four role-specific LLM agents (i.e., \textit{Analyst}, \textit{Modeler}, \textit{Designer}, and \textit{Evaluator}) to jointly derive, model, and evaluate software architectures from given Software Requirements Specifications (SRS)~\cite{Bass2021SAP}. Our results show that MAAD distinguishes itself as a knowledge-driven architecture design methodology, and it outperforms general MAS like MetaGPT~\cite{hong2023metagpt}, particularly in terms of architectural completeness. Moreover, MAAD seamlessly integrates external knowledge sources, supporting the addition of authoritative literature and private knowledge bases for domain-specific best practices. This extensibility enables MAAD to tailor architecture generation to specialized domains, mitigate hallucinations through logical reasoning, and deliver consistent, high-quality architecture designs with minimal human intervention. By further comparing three LLMs (i.e., GPT-4o, DeepSeek-R1, and Llama 3.3) as the foundational LLMs supporting the agents, the results show that the MAAD framework equipped with GPT-4o can achieve relatively better performance than the other two LLMs in our study. By pioneering a domain-aware, automated approach to architecture design, MAAD lays the foundation for next-generation development platforms that deliver rapid, reliable, and maintainable software architectures with minimal human oversight. Our \textbf{contributions} are threefold:

\begin{itemize}
    \item \textbf{Framework Design}: We present the architecture of MAAD and detail inter-agent protocols for collaborative reasoning. The four agents of MAAD can mirror the realistic architecture design process and generate complete architecture designs with evaluation reports.
    \item \textbf{Knowledge Integration}: We demonstrate how MAAD ingests and applies external knowledge sources to ground architectural decisions and mitigate hallucinations. MAAD supports the automation of architecture design based on customized domain knowledge.
    \item \textbf{Empirical Evaluation}: Through quantitative and qualitative assessments, we show that MAAD yields better architecture designs than the existing MAS baseline MetaGPT~\cite{hong2023metagpt} and significantly reduces manual intervention.
\end{itemize}

This paper is an extension of our previously published vision paper~\cite{Zhang_MAAD2025} by delivering the following substantial contributions. In comparison, this extended version (1) introduces the detailed description of each agent's implementation and their interaction mechanism, (2) evaluates the MAAD's performance and compares it with a baseline MetaGPT~\cite{hong2023metagpt}, (3) explores the influence of incorporating external knowledge on architecture design based on the same requirements specifications, (4) investigates the performance differences among three leading foundational LLMs (i.e., GPT-4o, DeepSeek-R1, and Llama 3.3), (5) validate the MAAD's practical effectiveness with the challenges and suggestions for future improvement through interviews with three architects, and (6) analyzes the advantages of MAAD over MetaGPT with a discussion of the impact of infusing external knowledge.

The remainder of this paper is organized as follows: Section~\ref{sec:RelatedWork} introduces related studies of this work. Section~\ref{sec:MAADFramework} elaborates on the design of the MAAD framework, and Section~\ref{sec:StudyDesign} presents the research questions. Section~\ref{sec:Results} describes the study results and our findings, and Section~\ref{sec:Discussions_and_Implications} discusses the results of this study. Section~\ref{sec:Threats} examines the threats to the validity of this study. Section~\ref{sec:Conclusion} summarizes this study and outlines the future work.

\section{Related Work}\label{sec:RelatedWork}

\subsection{Software Architecture Design}

Software architecture design has evolved significantly over the past few decades, transitioning from experience‑based heuristics to systematic, model‑driven engineering practices~\cite{Bass2021SAP}. Early approaches in the 1990s emphasized layered architectures and object-oriented decomposition, which were mainly guided by expert judgment and best practices~\cite{Shaw1996sap}. These foundational efforts defined architecture as a high-level abstraction encompassing system structure, behavior, and key quality attributes~\cite{Wan2003SAP}. As systems grew in complexity, researchers introduced Architecture Description Languages (ADLs) to bring formality to architectural modeling and analysis. Prominent ADLs such as AADL~\cite{Feiler2006AADL} allowed researchers and practitioners to define system components, connectors, and configurations systematically~\cite{Medvidovic2000ADL}. 

With the rise of distributed and service-oriented computing in the 2000s, Service-Oriented Architecture (SOA) became a dominant paradigm. SOA promoted loose coupling and service abstraction, facilitating scalability and adaptability in enterprise environments. Around the same time, Model-Driven Architecture (MDA) further emphasized the transformation of abstract architectural models into platform-specific implementations through model transformations, bridging the gap between abstract design and executable systems~\cite{Mellor2004MDAD}.

In the past decade, the rise of cloud-native systems, microservices, and event-driven architectures has reshaped software architecture around modularity, scalability, and deployment agility. These trends were accompanied by tools like Kubernetes, practices such as Domain-Driven Design (DDD)~\cite{Evans2004DDD}, and architectural styles like serverless computing and containerization. In parallel, architectural decision modeling and quality attribute-driven design approaches like Architecture Tradeoff Analysis Method (ATAM)~\cite{Kazman2000ATAM} have provided systematic ways to align architectural choices with business goals and non-functional requirements~\cite{Bass2021SAP}. 

Looking forward, the integration of Artificial Intelligence (AI) into architecture design is emerging as a transformative force~\cite{ahmad2023towards}. Knowledge‑based systems and AI‑assisted tooling are beginning to automate routine architectural decisions, generate candidate designs, and adapt architectures in response to evolving requirements~\cite{Espositoa2025GenAI4SA}.

Overall, the software architecture landscape is undergoing a transformation driven by AI. Architectural practices are increasingly integrating intelligent tooling and AI-based components capable of learning, adapting, or generating architecture elements. The modern architecture design process is becoming more iterative, model-centric, quality-aware, and AI-powered, supported by both theoretical frameworks and practical intelligent tools.

\subsection{Large Language Models for Software Architecture Design}

Large Language Models (LLMs) have received significant attention from both academia and industry community due to their remarkable performance across a wide range of Software Engineering (SE) tasks \cite{li2025ChatGPT}. Recent studies have begun to investigate the intersection between LLMs and software architecture design, highlighting the potential of LLMs to enhance architecture design processes and decision-making.

Schmid \textit{et al}.~\cite{Schmid2025LLMSA} conducted a systematic literature review by analyzing 18 studies on the application of LLMs in architectural tasks (e.g., design‑decision classification, pattern detection). Their work identifies emerging use of LLM techniques but also highlights underexplored areas (e.g., code‑generation from architecture and architecture conformance checking) and calls for stronger architecture evaluation frameworks. 
Esposito \textit{et al}.~\cite{Espositoa2025GenAI4SA} conducted a multivocal literature review synthesizing 37 sources, including both academic and gray literature. They identified key challenges regarding the use of LLMs for architecture design, such as LLMs' accuracy issues, hallucinations, ethical and privacy concerns, the absence of architecture‑specific datasets, and a dearth of architecture evaluation frameworks. Moreover, they advocated for research into general architecture evaluation methodologies, LLMs' transparency and explainability, ethical guidelines, and tailored benchmarks to support real‑world adoption. 
Eisenreich \textit{et al}.~\cite{Eisenreich_2024} proposed a semi-automated approach for generating candidate software architectures directly from requirements using LLMs. Their work demonstrates the feasibility of leveraging natural language requirements to guide early-stage architectural decision-making. 
Dhar \textit{et al}.~\cite{dhar2024can} examined the use of LLMs for generating Architecture Decision Records. Their study found that while GPT-4 can generate relevant design decisions in zero-shot settings, its performance does not yet match human-level reasoning. Interestingly, their results suggest that more cost-efficient models, such as GPT-3.5, can reach competitive performance under few-shot settings.

Despite these promising advancements, the application of LLM-based agents, particularly multi-agent systems, to software architecture design remains relatively underexplored compared to their use in other SE activities~\cite{Espositoa2025GenAI4SA, He2025mas}. Current research primarily focuses on single-agent reasoning or generation tasks regarding certain architecture activities, resulting in a notable gap in understanding of how distributed or collaborative LLM agents might co-design, evaluate, and iteratively refine software architecture in a more autonomous or interactive way.

\section{MAAD Framework}\label{sec:MAADFramework}

In this section, we present our proposed MAAD framework in three parts: first, we provide an overview of the MAAD framework (see Section~\ref{sec:Overview}); second, we detail the design and specifications of the constituent agents of the MAAD framework (see Section~\ref{sec:AgentDesign}); and third, we describe the collaborative mechanisms of the MAAD framework (see Section~\ref{sec:AgentCollaboration}). 

\subsection{Overview}\label{sec:Overview}

\begin{figure}[hbtp]
    \centering
    \includegraphics[width=\linewidth]{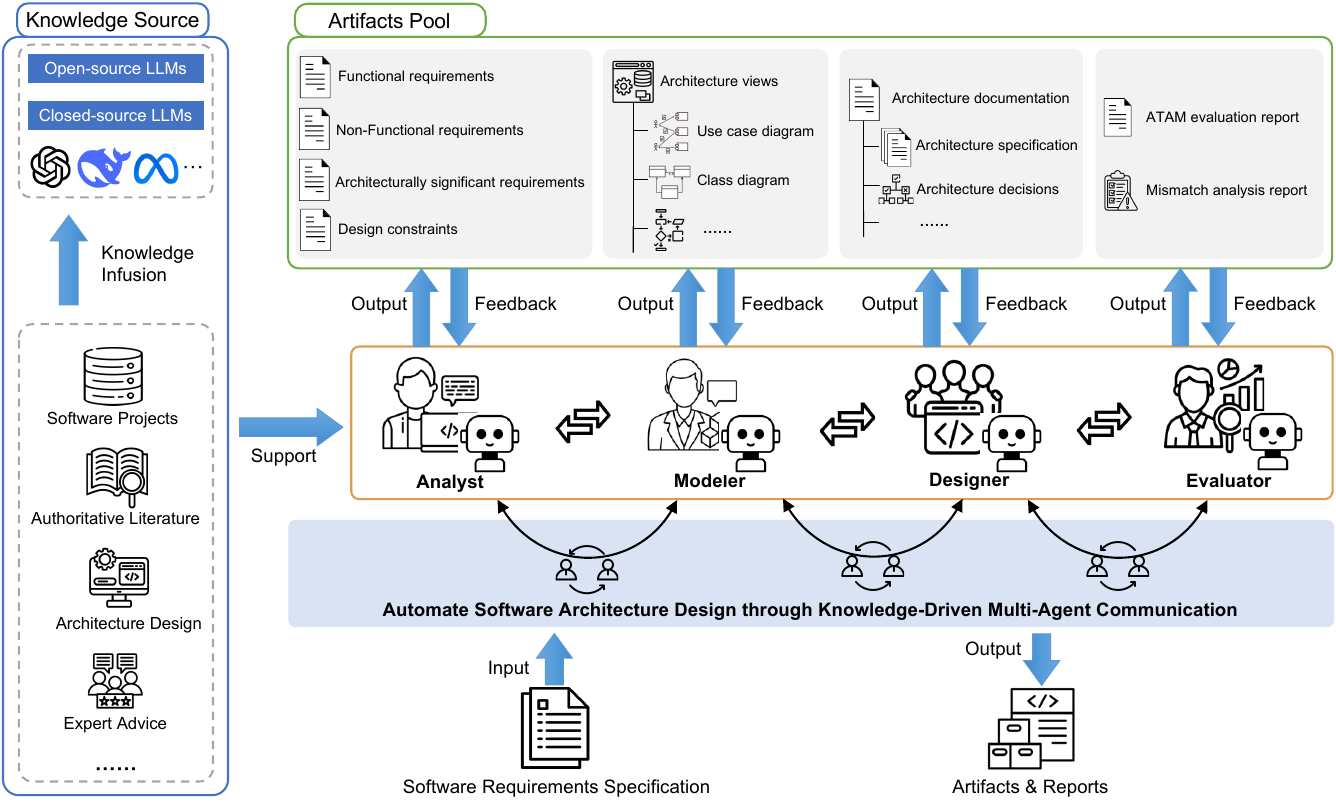}
    \caption{Overview of the MAAD framework}\label{F:Overview}
\end{figure}

The MAAD framework implements a knowledge‑driven, multi‑agent pipeline that autonomously transforms a Software Requirements Specification (SRS) into a complete architecture design (see Figure~\ref{F:Overview}). MAAD comprises four specialized agents, and each individual agent is equipped with perception, reasoning, and action capabilities. 

As illustrated in Figure~\ref{F:Overview}, after receiving the SRS, the \textbf{Analyst agent} examines its content, identifies and distills key aspects of the requirements into four decomposed requirements artifacts. Once the \textbf{Modeler agent} perceives the generated artifacts in the artifacts pools from the \textit{Analyst} agent, it constructs the system's multi‑view representation according to the ``4+1'' architecture view models proposed by Philippe Kruchten \cite{Kruchten1995avm}, which is a widely used architecture description framework in practice. When the \textbf{Designer agent} perceives the artifacts by the \textit{Analyst} and \textit{Modeler} agents, it synthesizes the final architecture documentation. The documentation formulates system goals and detailed design specifications. Finally, the \textbf{Evaluator agent} assesses the generated architecture by checking the architecture views against the original SRSs. It produces an ATAM Evaluation report and a Mismatch Analysis report, pinpointing any deviations between generated artifacts and SRSs, and trade‑offs during architecture design. Overall, the four agents enact a cohesive and feedback‑driven workflow. Through iterative communication and artifact exchange, MAAD ensures that each agent's autonomous activities cumulatively yield a robust, traceable architecture design.

\subsection{Agent Design}\label{sec:AgentDesign}
\subsubsection{Analyst Agent}\label{sec:AnalystAgent}

The \textit{Analyst} agent serves as a critical component responsible for comprehending the SRS, filtering and classifying requirements, and documenting those requirements that exert significant influence on the architectural design. This agent plays a pivotal role in ensuring alignment between subsequent architectural decisions and overarching business objectives. The \textit{Analyst} agent is designed to produce four decomposed requirements artifacts: \textit{Functional Requirements}, \textit{Non-Functional Requirements}, \textit{Architecturally Significant Requirements}, and \textit{Design Constraints}.
\begin{comment}
\begin{itemize}
    \item \textit{Functional Requirements} refer to the system's anticipated behaviors and operational capabilities.
    \item \textit{Non-Functional Requirements} establish performance criteria and quality attributes, such as scalability, reliability, and security.
    \item \textit{Architecturally Significant Requirements (ASRs)} identify and prioritize requirements that exert substantial influence on architectural decisions, including modularity or integration constraints.
    \item \textit{Design Constraints} outline limitations and restrictions that must be observed throughout the design process, including technological boundaries and regulatory compliance requirements.
\end{itemize}
\end{comment}

To accomplish this task, the \textit{Analyst} agent possesses capabilities to comprehend the SRS through the following systematic actions: (1) parsing and structuring the SRS to extract all requirements comprehensively; (2) identifying and filtering Architecturally Significant Requirements (ASRs); (3) classifying requirements into functional and non-functional categories, where functional requirements specify system features and operational tasks, while non-functional requirements encompass quality attributes, resource constraints, and other pertinent considerations; and (4) extracting system design constraints from the SRS.
\begin{comment}
For instance, the prompt design employed for ASRs extraction by the Analyst agent is defined in Figure~\ref{fig:prompt4analyst_ASR}.

% 移除prompt之后，增加prompt设计的说明
\begin{figure}[h]
    \begin{tcolorbox}[sharp corners, width=\textwidth]
    \footnotesize
    Given the software requirements below\\
        \{requirement\_documents\}\\
        Generate Architecturally Significant Requirements (ASRs) from the provided requirements.
        For each ASR, the criteria are: influence system architectural decisions, involve critical quality attributes, require cross-component coordination.\\
        Structure your outputs as follows:\\
        ``The ASRs are:\\
        1. [ASR-001]:\\
        - [Original text of ASR1]\\
        - [Related quality attribute(s)]\\
        - [Architectural Impact]\\
        2. [ASR-002]:
        ...''\\
        An example is here:\\
        1. ASR-001\\
        - Original text of ASR1: The system must support 100,000 concurrent requests per second
        - [Related quality attribute(s)]: Scalability, Availability, Performance efficiency, Fault Tolerance\\
        - Architectural Impact: A distributed architecture design is required\\
        - Related Components: API Gateway, Load Balancer
    \end{tcolorbox}
    \caption{Prompt of the Analyst agent to generate architecturally significant requirements (ASR)}
    \label{fig:prompt4analyst_ASR}
\end{figure}
\end{comment}

\subsubsection{Modeler Agent}\label{sec:ModelerAgent}

The \textit{Modeler} agent aims to translate the requirements into a software architecture blueprint, specifically to model the overall architecture of the system based on the refined requirements provided by the \textit{Analyst} agent. This agent generates the ``4+1'' architecture view models \cite{Kruchten1995avm} as its generated artifacts, including \textit{Logical Views}, \textit{Development View}, \textit{Process View}, \textit{Physical View}, and \textit{Scenario View}.

To fulfill these responsibilities, the \textit{Modeler} agent executes the following actions: (1) prioritizing non-functional requirements to facilitate strategic trade-offs among competing quality attributes; (2) selecting appropriate technology stacks that align with system requirements and constraints; (3) identifying and applying suitable architectural styles and patterns to structure the system effectively; and (4) analyzing and modeling domain components along with their interrelationships to construct comprehensive architecture views.

\subsubsection{Designer Agent}\label{sec:DesignerAgent}

The \textit{Designer} agent is in charge of generating detailed architecture documentation based on the architecture views from the \textit{Modeler} agent, thereby establishing a robust foundation for subsequent code implementation. This documentation encompasses the following components:

\begin{itemize}
    \item \textit{Goals}: Define the primary objectives and quality attributes that the proposed architecture aims to achieve, such as performance targets, scalability requirements, and maintainability.
    \item \textit{Detailed Architecture Design}: A detailed design specification that defines the system's components, modules, and subsystems while outlining their interactions, interfaces, and behavioral contracts.
    \item \textit{Component \& Connector Specifications}: Detailed descriptions of communication protocols (e.g., REST API contracts, event-driven messaging formats, database connection specifications) with explicit error-handling mechanisms, performance thresholds, and reliability constraints.
    \item \textit{Key Technologies}: Comprehensive infrastructure specifications, such as resource allocation strategies, deployment configurations, and explicit analysis of scalability versus fault-tolerance trade-offs.
    \item \textit{Design Decisions}: Systematic explanation of architectural decisions, encompassing selected architecture patterns, technology choices underlying key structural and behavioral design choices.
    \item \textit{Design Decision Rationale}: In-depth justification of technology selections, pattern adoption, and architectural trade-off analysis with consideration of alternative approaches and their respective limitations.
    \item \textit{Executable Prototype Skeleton}: Generation of structured code scaffolding and skeletal implementations for critical system modules, including interface and basic operational logic.
\end{itemize}

To fulfill its responsibilities, the \textit{Designer} agent executes the following actions: (1) analyzing and interpreting architecture views to extract design requirements and constraints; (2) synthesizing comprehensive design specifications that bridge conceptual models with implementable solutions; (3) defining precise component interfaces and interaction protocols to ensure seamless system integration; (4) documenting architectural decisions with explicit rationale and trade-off analysis to support future maintenance and evolution; and (5) generating code skeleton that provides structured foundations for development teams. 

\subsubsection{Evaluator Agent}\label{sec:EvaluatorAgent}

The \textit{Evaluator} agent is tasked with rigorously assessing the architectural artifacts generated by other agents to ensure their alignment with the input SRS. This agent produces two main evaluation reports:

\begin{itemize}
    \item \textit{ATAM Evaluation Report}: A comprehensive architectural assessment based on the Architecture Tradeoff Analysis Method (ATAM), systematically evaluating quality attribute scenarios, identifying architectural strengths and weaknesses, analyzing potential risks, and documenting trade-off implications across competing quality attributes.
    \item \textit{Mismatch Analysis Report}: A detailed diagnostic analysis that identifies and categorizes the discrepancies between the generated architectural solutions and the original requirements specifications, providing critical insight into areas requiring architecture refinement, requirements clarification, or design iteration.
\end{itemize}

To fulfill these responsibilities, the \textit{Evaluator} agent performs the following actions: (1) conducting a comprehensive evaluation of generated architectural artifacts, including structural diagrams (class, component, package), behavioral diagrams (sequence, activity, state), and deployment specifications to identify potential risks with ASRs and design constraints as specified in the SRS; and (2) performing systematic mismatch analysis through requirements traceability matrices to quantify and categorize identified discrepancies.%; (3) generating actionable recommendations for architectural refinement based on evaluation findings and trade-off analysis. 
\begin{comment}
The prompt design employed for producing two architecture evaluation reports by the Evaluator agent is demonstrated in Figure~\ref{fig:prompt4evaluator_MR}.

\begin{figure}[h]
    \begin{tcolorbox}[sharp corners, width=\textwidth]
    \footnotesize
    According to the provided requirements and UML diagrams (recorded in PlantUML syntax), please generate a Mismatch Analysis Report. \\
    1. Requirement: \{requirement\_document\}\\
    2. Architecture views: \{AV\}\\
    3. Architecture documentation: \{AD\}\\
    Your report should identify any discrepancies, inconsistencies, or gaps between the system requirements and the architectural design. Structure your outputs as follows:\\
    ``[mismatch 1]\\
    - Description: A brief explanation of the mismatch.\\
    - Impact: An analysis of the potential impact on the system.\\
    - Recommendation: Suggestions or steps to resolve or mitigate the issue. \\ 
   
    [mismatch 2]\\
    - Description: A brief explanation of the mismatch.\\
    - Impact: An analysis of the potential impact on the system.\\
    - Recommendation: Suggestions or steps to resolve or mitigate the issue.\\
    ……''
    \end{tcolorbox}
    \caption{Prompt of the Evaluator agent to generate mismatch report (MR)}
    \label{fig:prompt4evaluator_MR}
\end{figure}
\end{comment}

\subsubsection{Knowledge-driven Agent Setting}\label{sec:AgentSetting}

% 【知识注入的方法】知识图谱注入 (Knowledge Graph Infusion)、知识蒸馏 (Knowledge Distillation)、外部知识嵌入 (External Knowledge Embeddings)、多模态学习 (Multimodal Learning)、增强训练数据 (Augmented Training Data)、记忆增强模型 (Memory-Augmented Models)、符号推理与神经网络的结合 (Symbolic-Reasoning with Neural Networks)、任务定制知识注入 (Task-specific Knowledge Infusion)

% 【本文使用的知识注入的方法】外部知识嵌入 (External Knowledge Embeddings)，将教材作为external knowledge作为embedding嵌入
% 只使用教材到RAG的话，可解释性存在问题；需要展示可用性
% 不同场景、领域的软件，架构设计都不一样；
% 这个讨论部分补充

In a multi-agent framework like MAAD, the infusion of external knowledge is essential to equip agents with the insights they need for high‑quality, context‑aware artifact generation. Software architecture design is inherently complex: agents cannot rely solely on initial requirements and intermediate artifacts but also leverage existing, external knowledge, ranging from industry standards, architecture patterns, and domain‑specific best practices. By integrating external knowledge, agents can reduce informational gaps, mitigate the risk of omitting critical considerations, and ensure that the generated artifacts reflect both theoretical principles and empirical insights.

% \yiran{I think we need more details here? 1) Why only modeler and designer infused with knowledge? 2) What is the exact knowledge source? 3) How is the database segmented? 4) What are the inputs for the vector search and what algorithm is applied to search the knowledge?}

In our MAAD framework, we focus knowledge infusion on the two agents whose tasks most critically depend on external context:

\textbf{Modeler Agent}: Once receiving the artifacts generated by the \textit{Analyst} agent, the \textit{Modeler} agent enhances its understanding by performing a similarity search over a vectorized knowledge base. This database is segmented into thematic categories (e.g., layered architecture, component‑and‑connector style), and the segments are then incorporated into the prompt fed to the \textit{Modeler} agent. The integration of this external knowledge allows the \textit{Modeler} agent to incorporate broader perspectives on architecture design, ensuring that the architecture views it generates are grounded in both the provided requirements artifacts and additional contextually relevant information.

\textbf{Designer Agent}: Similarly, before generating the architecture documentation, the \textit{Designer} agent performs a vector search in the same external knowledge base. It retrieves the three most pertinent text segments, which are then used as background knowledge to inform the architecture design process. By referencing these external insights, the \textit{Designer} agent can ensure that the design rationale and decisions align with industry standards and proven architectural patterns, further refining the quality of the generated architecture documentation.

Through pilot experiments, we specifically chose to retrieve the three most similar text segments to the prompt provided to the agents. This decision was driven by the need to strike a balance between relevance and conciseness for the agents' generated artifacts. By selecting only the most relevant segments, we ensure that agents receive targeted knowledge without overwhelming them with excessive information, which could lead to unnecessary verbosity or potential confusion in the generated artifacts.

This approach of infusing external knowledge not only enhances the agents' decision-making but also ensures that the generated artifacts are aligned with established best practices and external expertise, ultimately contributing to the robustness and effectiveness of the architectural design process within the MAAD framework.

\subsection{Agent Collaboration}\label{sec:AgentCollaboration}
% Initially, the Analyst agent analyzes the input SRS by extracting requirements and constraints that influence the architecture design. 
% Building on the Analyst agent's analysis results, the Modeler agent undertakes two primary tasks: 1) formulating high-level architectural decisions to guide the design, and 2) specifying the domains and prioritized quality attributes that the generated system needs to address. Next, the Designer agent conducts concrete architecture implementation by creating UML diagrams (class diagrams, sequence diagrams, and deployment diagrams, etc.), which serve as blueprints to guide follow-up code development. 
% Finally, the Evaluator agent verifies whether the artifacts align with the input SRS. If any mismatches are found, it identifies the root cause and collaborates with the relevant agent to resolve the issue. The remaining agents then update their respective artifacts accordingly. 
% The architecture design process concludes when the Evaluator confirms the artifacts. The final architecture design, including the designed diagrams, conceptual views, and documented architectural decisions, collectively serve as the foundation for the subsequent code implementation.

The architecture design process of the MAAD framework begins with the \textbf{Analyst agent}, which meticulously analyzes the input Software Requirements Specification (SRS). The agent extracts key requirements and constraints that are critical to the architecture design, ensuring a thorough understanding of the SRSs that will guide subsequent steps.

Building on the results of the \textit{Analyst} agent's analysis, the \textbf{Modeler agent} takes on two primary responsibilities: 1) formulating high-level architectural decisions that establish a clear direction for the system's design, and 2) identifying the relevant domains and prioritizing the quality attributes (e.g., performance, security, scalability) that the generated system must address. These tasks ensure that the architecture is both robust and aligned with the functional and non-functional requirements outlined in the SRS.

Following the high-level decisions made by the \textit{Modeler} agent, the \textbf{Designer agent} focuses on the concrete implementation of the architecture. The \textit{Designer} agent creates detailed Unified Modeling Language (UML) diagrams, including class diagrams, sequence diagrams, and deployment diagrams that serve as comprehensive blueprints for the system's design. These diagrams not only capture the structure and interactions of the system, but also provide an essential reference for guiding the subsequent code development, ensuring alignment between design and implementation.

Then, the \textbf{Evaluator agent} verifies the consistency and alignment of the generated architecture artifacts with the original SRS. If mismatches are identified, the \textit{Evaluator} agent works to pinpoint the root cause of the discrepancies. It collaborates with the relevant agents (the \textit{Analyst}, \textit{Modeler}, or \textit{Designer} agents) to resolve the issue. Following the resolution, the affected agents update their respective artifacts, ensuring that the entire architecture design is coherent and consistent with the SRS.

Once the \textit{Evaluator} agent confirms that all artifacts align with the SRS and that all issues have been addressed, the architecture design process concludes. The final architecture design, which includes detailed UML diagrams, conceptual views, and documented architectural decisions, forms the foundation for subsequent code implementation. This structured approach ensures that the system's architecture is both comprehensive and well-aligned with the original SRSs, providing a solid basis for efficient and effective software development.

\section{Study Design}\label{sec:StudyDesign}
In this section, we present the Research Questions (RQs) in Section~\ref{sec:RQs} and provide an overview of the research process in Section~\ref{sec:Experiment Settings}.

\subsection{Research Questions}\label{sec:RQs}
% \subsection{RQ1: }\label{sec:RQ1_results}

\begin{tcolorbox}[colback=gray!8, colframe=gray]
% \textbf{RQ1: How can an LLM-based multi-agent framework automate the architecture design?}
\textbf{RQ1: How effective is MAAD in automating the software architecture design?}
\end{tcolorbox}

\textbf{Rationale}: MAAD employs a multi-agent framework to automate the software architecture design. While such automation presents considerable potential, it is essential to evaluate its practical effectiveness. Therefore, RQ1 seeks to evaluate the effectiveness of MAAD in generating viable and relevant software architectures. To achieve a comprehensive evaluation, we will first perform a comparative analysis of MAAD's performance against baseline methods. Then we will present a case study where MAAD is applied to a realistic architecture design scenario to illustrate its strengths and weaknesses. Moreover, we conduct a human evaluation of the generated architecture with industry architecture experts to gain more insights on MAAD's performance.
By answering this question via this multifaceted evaluation, we aim to provide empirical evidence of MAAD's capability to effectively automate architecture design, benchmark its performance, and validate its potential to enhance efficiency and scalability in real-world software development contexts.

% \textbf{Rationale}: The architecture design traditionally requires significant human expertise, collaboration, and manual effort. However, leveraging LLMs within a multi-agent framework presents an opportunity to automate this complex process. RQ1 aims to explore how LLM-based agents can be used to automate the architecture design. This process mimics the decision-making, problem-solving, and design processes involved in architecture development. By answering this question, we aim to clarify the process and details through which automation can streamline the architecture design, improving efficiency and scalability in software development projects.
% 结果：拿1个需求文档的case展示一下每个agent输出的artifacts（图）
% 对比MAAD跟metagpt（这种通用框架）生成的制品做对比
% 定量：计算需求--架构的mismtach；
% 定性：ATAM方法
% 【改动】把RQ1的内容合并到Sec 3.2～3.5；RQ4的interview作为RQ1的一部分
% RQ1的结果展示：case study展示每个agent生成的制品；（2）对比MetaGPT的生成制品的结果；（3）interview的结果，证明MAAD框架的有效性，RQ4就取消掉。

\begin{tcolorbox}[colback=gray!8, colframe=gray]
\textbf{RQ2: To what extent can infusing external knowledge help improve the quality of architecture design in the MAAD framework?}
\end{tcolorbox}
\textbf{Rationale}: The MAAD framework can leverage the infusion of external knowledge (e.g., knowledge from existing system designs, authoritative literature, and architecture experts) to improve trustworthiness and correctness. The integration of external knowledge is crucial to ensure that agents in the MAAD framework can make informed decisions that align with best practices and real-world requirements. RQ2 investigates methods for infusing and synthesizing this knowledge, such as knowledge extraction and knowledge representation frameworks. Addressing this RQ will help improve the accuracy and effectiveness of the artifacts generated by the agents, ensuring that the architecture aligns with both theoretical foundations and practical constraints.
% 结果：对比加和不加external knowledge时，LLM生成制品的区别？体现出知识注入对架构设计的影响（目前没有相关工作）

\begin{tcolorbox}[colback=gray!8, colframe=gray]
\textbf{RQ3: How do different LLMs affect the quality of architecture design in the MAAD framework?}
\end{tcolorbox}
\textbf{Rationale}: Different LLMs vary in terms of their generations of underlying architecture views and specialization, which may influence their ability to generate high-quality software architecture designs. RQ3 investigates the impact of various LLMs, such as GPT-4o, DeepSeek R1, and Llama 3.3 within the MAAD framework. By understanding which LLMs are most suitable for specific architecture tasks (e.g., requirements analysis, trade-off evaluations), we can refine the framework to ensure optimal performance and reliability in automated software architecture design.
% 结果：gpt-4o，llama 3.3，deepseek-r1，对比不同LLM在同一套prompt下，输出架构制品的差异(比较生成views/diagrams的语法树的差异)；评估的话，就对比evalutor agent的结果

\begin{comment}
\begin{tcolorbox}[colback=gray!8, colframe=gray]
\textbf{RQ4: What are the main challenges encountered in deploying LLM-powered agents for automated architecture design?}
\end{tcolorbox}
\textbf{Rationale}: While LLM-powered agents offer great potential for automating architecture design, there are several challenges in their deployment. These challenges may include knowledge gaps, model biases, limited adaptability to new requirements, and issues related to the integration of agent generations into coherent architecture solutions. Additionally, practical deployment concerns such as scalability, real-time feedback, and maintaining the quality of generated designs over time need to be addressed. RQ4 aims to identify these challenges and propose solutions or strategies to mitigate them through interviews with sophisticated architects. Understanding these obstacles is essential for refining the MAAD framework and ensuring its robustness and practical applicability in real-world software development environments.

% 结果：找架构师做个interviews，从practitioners的经验谈一下自动化架构设计面临的挑战
\end{comment}

\subsection{Experiment Settings}\label{sec:Experiment Settings}

\begin{figure}[hbtp]
    \centering
    \includegraphics[width=0.9\linewidth]{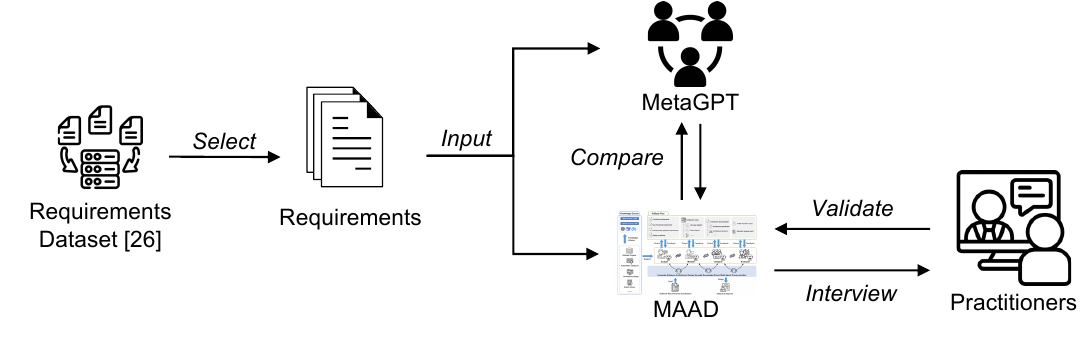}
    \caption{Process of the sudy design}\label{F:StudyDesign}
\end{figure}

This section outlines the experimental design used to validate our proposed Multi-Agent Architecture for Development (MAAD) framework. Figure~\ref{F:StudyDesign} presents the study design of our work, which includes requirements dataset selection, comparison between MAAD and a baseline (i.e., MetaGPT~\cite{hong2023metagpt}), and validation from practitioners (architects).

\subsubsection{Dataset}\label{sec:Dataset}
The requirements dataset is collected from Jin \textit{et al.} \cite{Jin2024erm}, which was gathered from public datasets, including PURE~\cite{Ferrari2017Pure} (a dataset of 79 publicly available natural language requirements documents collected from the Web) and private industrial requirements documents~\cite{yang2022ipsf}. We then select the requirements of the ``Space Fraction System'' (SFS) as the input to conduct a case study in this work.

\subsubsection{Baseline}\label{sec:Baseline}
To compare our proposed MAAD framework with peer MAS, we select MetaGPT \cite{hong2023metagpt} as the baseline. MetaGPT is a state-of-the-art multi-agent system that generates comprehensive software artifacts from a single requirements statement. Structurally, MetaGPT encompasses four agents \textit{Product Managers}, \textit{Architects}, \textit{Project Managers} and \textit{Engineers}. MetaGPT provides the entire process of a software company along with carefully orchestrated Standard Operating Procedures (SOPs) \cite{hong2023metagpt}. 

\subsubsection{Large Language Model Selection}\label{sec:LLMSelection}
To ensure comprehensive evaluation across diverse model configurations and capabilities, we select three representative high-performance LLMs as the foundational LLMs for MAAD based on their technical diversity, capabilities, and availability. 

\begin{itemize}
\item \textbf{GPT-4o}: OpenAI's proprietary multimodal LLM developed \cite{GPT4} with approximately 200 billion parameters. It excels at code generation and mathematical reasoning, with fast 320ms response times for real-time interactions, and it has been commonly used in business applications, education, and content creation.

\item \textbf{Llama 3.3}: Meta's latest open-source LLM (70 billion parameters) with an extended 32K-token context window \cite{dubey2024llama}. It improves multilingual understanding and generation, matches GPT-4 on many NLP benchmarks, and offers efficient fine-tuning for custom applications.

\item \textbf{DeepSeek-R1}: DeepSeek's open-source 671-billion-parameter model \cite{deepseekai2025} that cuts training costs by 60\% and boosts inference throughput by over 2.3× compared to dense 70B models. Its support for domain-specific tuning across specialized fields makes it relevant for evaluating adaptability in diverse SE contexts.
\end{itemize}

\subsubsection{Interviews}\label{sec:interview_design}

To further validate the correctness and practical utility of MAAD, we solicit feedback from experienced industry practitioners through semi-structured interviews.

\textbf{Interview Protocol}: We design an interview protocol by following the guidelines for empirical studies in software engineering proposed by Wohlin \textit{et al.}~\cite{Wohlin2012ESE}. We conduct semi-structured interviews with 3 open-ended questions that are designed to elicit practitioners' perspectives on automated architecture design using the MAAD framework. The interview questions allow the participants to freely and openly express their experiences and insights on the generated artifacts by MAAD from 11 real-world SRSs. The interview procedure consists of three parts: first, participants receive a concise overview of the study's objectives and are asked to review the artifacts generated by our framework for representative user requirements cases. Second, interviewees are asked demographic questions (e.g., role, years of professional experience). Third, the first author conducts the interviews, each of which lasted 35 to 50 minutes. With the interviewees' consent, we audio-record the interviews and fully transcribed them for an in-depth analysis. The interview protocol and open questions are available in our replication package~\cite{onlinepackage_TOSEM}.

\textbf{Data Analysis}: The first author conducts a qualitative analysis of the transcripts, with the fourth author independently reviewing all coded segments to ensure consistency and mitigate bias. The data analysis proceeds as follows: (1) Extracting data: Transcripts are carefully read to identify salient comments regarding MAAD's artifact quality; (2) Coding data: Initial codes are generated to categorize participants' views on MAAD's generated artifacts. These codes guide subsequent thematic analysis; (3) Examination: To ensure analytical rigor, the fourth author independently reviews the coded data and resolves any discrepancies through discussion.

\section{Results}\label{sec:Results}
\subsection{Results of RQ1}\label{sec:RQ1_Results}
\subsubsection{Generated Artifacts of the MAAD Framework}\label{sec:RQ1_p1}

% To answer RQ1, we employed the MAAD approach to generate software architecture design using the Software Requirements Specifications (SRSs). We use requirements of the ``Space Fraction System'' (SFS) as a running example, which were gathered from a public dataset PURE~\cite{Ferrari2017Pure}. The corresponding SRS is collected from Jin \textit{et al.}'s publication~\cite{Jin2024erm}.

To answer RQ1, we employed the MAAD approach to generate software architecture design using the Software Requirements Specifications (SRSs). We use requirements of the ``Space Fraction System'' (SFS) as a running example from our dataset (see Section~\ref{sec:Dataset}). %Jin \textit{et al.} \cite{Jin2024erm}, which were gathered from public datasets (e.g., PURE~\cite{Ferrari2017Pure}) and private industrial requirements documents~\cite{yang2022ipsf}. 
SFS is an interactive web-based educational platform that targets sixth-grade students and gamifies fraction arithmetic by presenting questions, providing immediate feedback, and tracking study performance. Students can answer fraction-related arithmetic questions, receive immediate feedback, and track their scores.

Upon ingesting the SRSs of the SFS case, the \textit{Analyst} agent (see Section \ref{sec:AnalystAgent}) automatically produces structured documentation and stores the four sets of artifacts: architecturally significant requirements (ASRs), functional requirements, non-functional requirements, and design constraints. Each artifact type is generated based on a predefined prompt template in which the complete SRS text is embedded% (e.g., Figure \ref{fig:prompt4analyst_ASR} illustrates the ASR template)
. Subsequently, the remaining three agents generate the other artifacts as defined in Section~\ref{sec:MAADFramework}: the \textit{Modeler} agent produces the ``4+1'' architecture view models \cite{Kruchten1995avm}, which comprise various UML diagrams organized as follows. 

\begin{itemize}
    \item \textit{Logical View} includes 3 UML diagrams: Class Diagram, Object Diagram, and State Diagram.
    \item \textit{Development View} includes Package Diagram and Component Diagram.
    \item \textit{Process View} includes Activity Diagram, Sequence Diagram, and Collaboration Diagram.
    \item \textit{Physical View} includes Deployment Diagram and Container Diagram.
    \item \textit{Scenario View} is also known as Use Case Diagram.
\end{itemize}

The \textit{Designer} and \textit{Evaluator} agents then generate the relevant architecture documentation and architecture evaluation reports as defined in Section~\ref{sec:DesignerAgent} and Section~\ref{sec:EvaluatorAgent}. All architecture artifacts generated by the MAAD approach are publicly available in the replication package~\cite{onlinepackage_TOSEM}.

\subsubsection{Comparison between MAAD and MetaGPT}\label{sec:RQ1_p2}

To evaluate MAAD's performance against the baseline, we conducted a comparative analysis with MetaGPT \cite{hong2023metagpt} (see Section~\ref{sec:Baseline}). %, a state-of-the-art multi-agent system that generates comprehensive software artifacts from a single requirements statement. Structurally, MetaGPT encompasses four agents \textit{Product Managers}, \textit{Architects}, \textit{Project Managers}, \textit{Engineers}. MetaGPT provides the entire process of a software company along with carefully orchestrated Standard Operating Procedures (SOPs) \cite{hong2023metagpt}. 
Since MetaGPT was not specifically designed for architecture design tasks, we focused our comparison on the overlapping artifact types produced by both systems, including artifacts of requirements analysis, system design, and documentation. This way allowed for a meaningful comparison of the capabilities of MetaGPT and MAAD within the same types of generated artifacts. To compare the artifacts generated by MAAD and MetaGPT, we conducted experiments on the two MAS frameworks, respectively. 

In terms of \textbf{requirements analysis}, the \textit{Product Manager} agent of MetaGPT reads the input SRSs and generates a structured SRSs, including: 

\begin{itemize}
    \item \textit{Product Goals}: define concise statement of the system’s primary features.
    \item \textit{User Stories}: describe user usage scenarios and interaction processes.
    \item \textit{Competitive Analysis}: compares competitors' features of similar products, highlights their strengths and weaknesses, and provides feature optimization recommendations.
    \item \textit{Requirement Analysis}: refines functional and non-functional requirements (such as performance and compatibility).
    \item \textit{Requirements Pool}: includes a prioritized requirements list.
    \item \textit{UI Design Draft}: presents sketch layouts of UI design and shows basic interface design descriptions.
\end{itemize}

% \begin{figure}
%     \centering\includegraphics[width=0.75\linewidth]{Figures/MetaGPT_requirement_analysis.png}
%     \caption{The requirement analysis results of SFS generated by MetaGPT}
%     \label{fig:MetaGPT_requirments}
% \end{figure}

Given the characteristics of MetaGPT, which ``\textit{takes a one-line requirement as input and outputs user stories, competitive analysis, requirements, data structures, APIs, documents, etc.}''\footnote{https://github.com/FoundationAgents/MetaGPT/}, it will automatically ``complete'' certain unreal details based on LLM's text generation capability, for example, specify the specific demands of students and teachers (e.g., ``track progress'' and ``update questions independently'') of the SFS project. Regarding the results of requirements analysis, MetaGPT cannot categorize requirements but only produce five top priority requirements for SFS, such as labeled as P0 (i.e., core functional requirements) or P1 (i.e., secondary functional requirements). By contrast, MAAD's \textit{Analyst} agent produced 6 functional requirements categories encompassing 21 detailed requirements, alongside 11 non-functional requirements and 8 architecture-related requirements. This finer granularity and greater coverage yield a more robust foundation for downstream architecture modeling.

\begin{figure}[htbp]
    \centering
    % 第一个子图 (a)
    \begin{subfigure}[b]{0.45\textwidth}
        \centering
        \includegraphics[width=\textwidth]{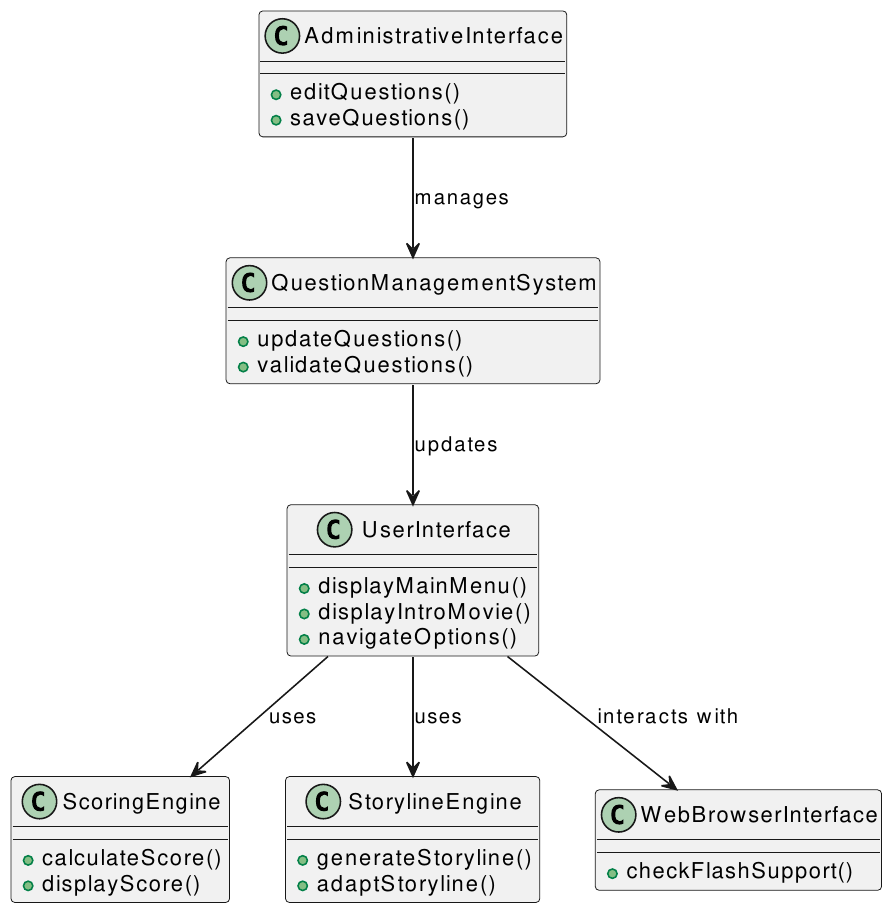} % 替换为MAAD生成的图片路径
        \caption{Class diagram generated by MAAD}
        \label{fig:CDmaad}
    \end{subfigure}
    \hspace{0.01\textwidth} % 宽度百分比控制 (推荐0.01-0.03)
    % 第二个子图 (b)
    \begin{subfigure}[b]{0.52\textwidth}
        \centering
        \includegraphics[width=\textwidth]{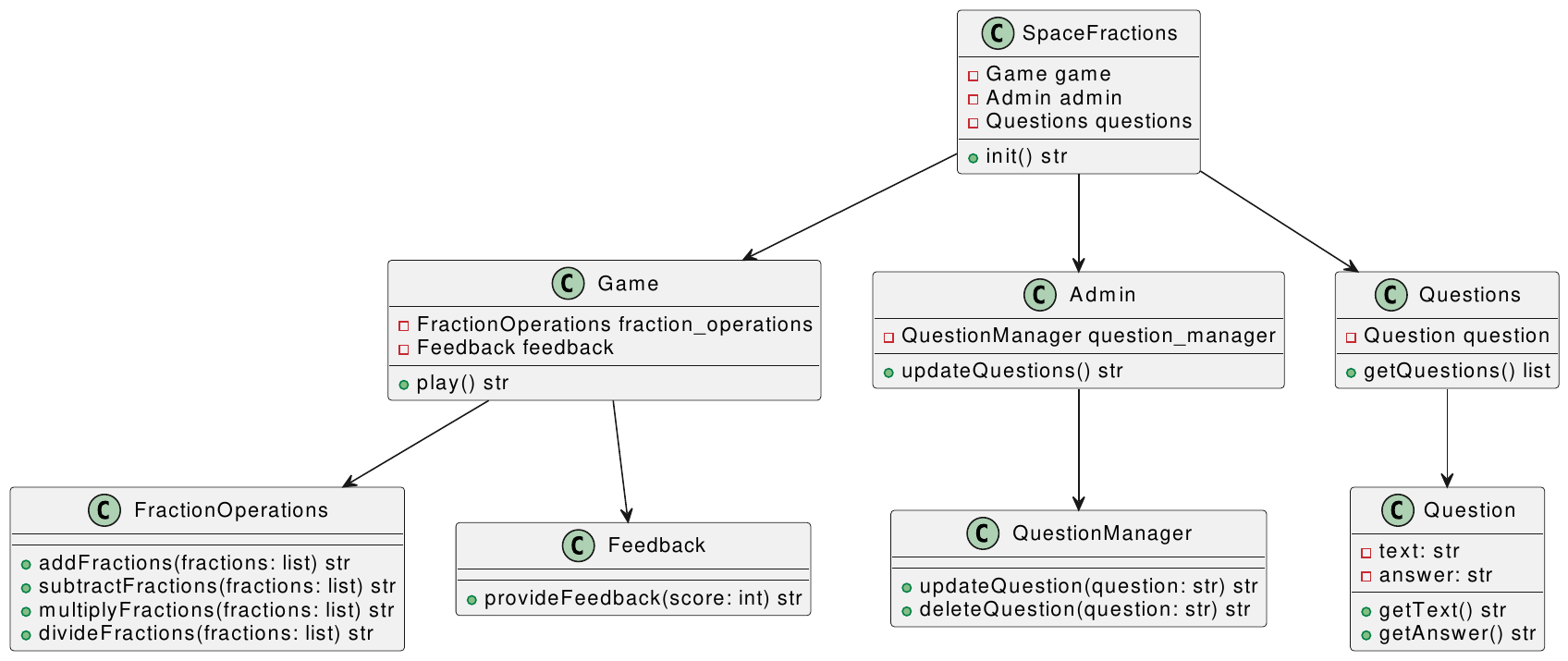}
        \caption{Class diagram generated by MetaGPT}
        \label{fig:CDmetagpt}
    \end{subfigure}
    \caption{Comparison of class diagrams generated by MAAD and MetaGPT}
    \label{fig:CDcomparison}
\end{figure}

\begin{figure}[htbp]
    \centering
    % 第一个子图 (a)
    \begin{subfigure}[b]{0.46\textwidth}
        \centering
        \includegraphics[width=\textwidth]{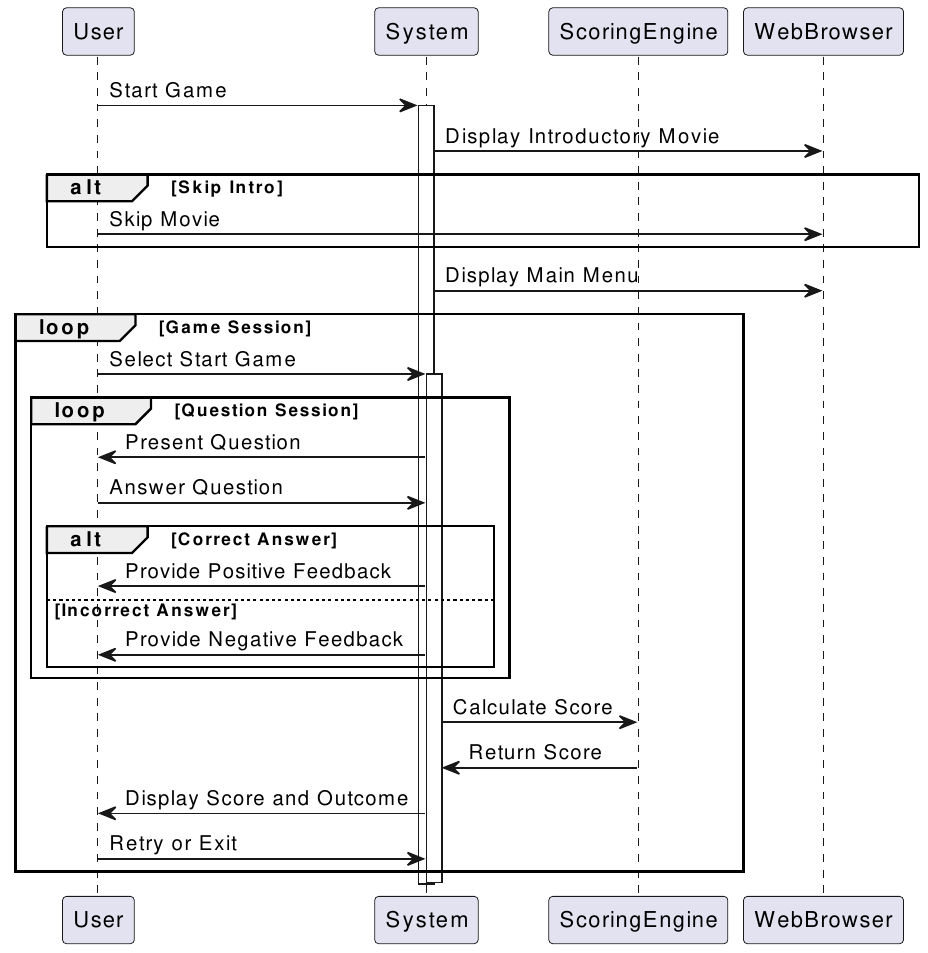}
        \caption{Sequence diagram generated by MAAD}
        \label{fig:SDmaad}
    \end{subfigure}
    \hspace{0.02\textwidth} % 宽度百分比控制 (推荐0.01-0.03)
    % 第二个子图 (b)
    \begin{subfigure}[b]{0.50\textwidth}
        \centering
        \includegraphics[width=\textwidth]{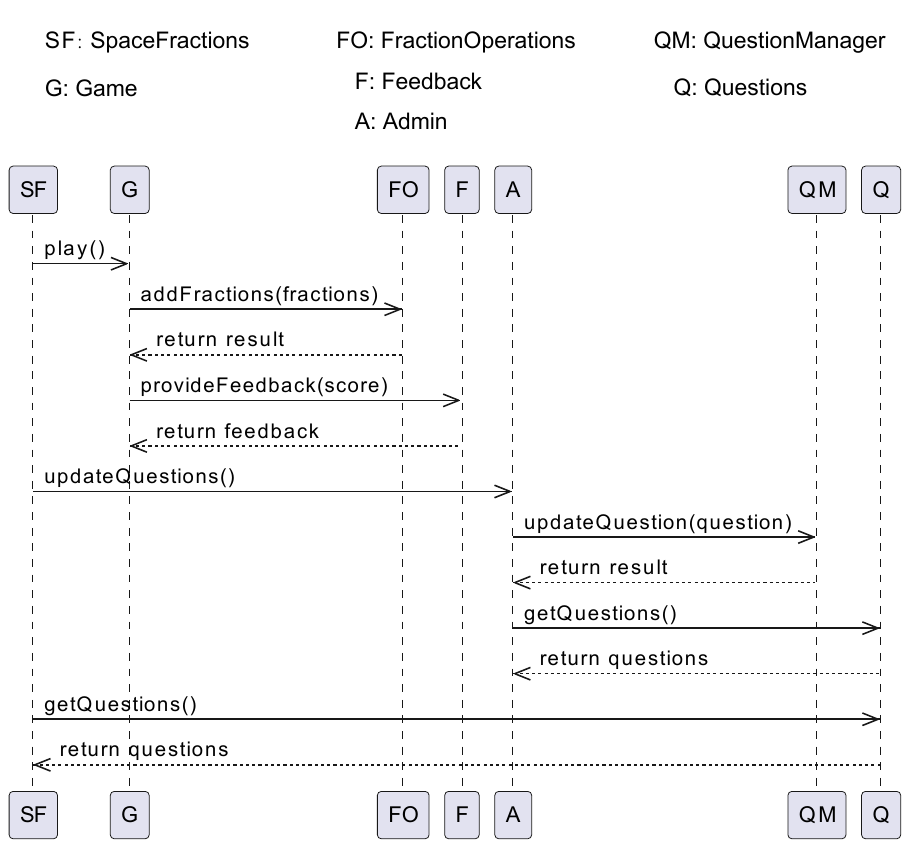}
        \caption{Sequence diagram generated by MetaGPT}
        \label{fig:SDmetagpt}
    \end{subfigure}
    \caption{Comparison of sequence diagrams generated by MAAD and MetaGPT}
    \label{fig:SDcomparison}
\end{figure}

Regarding \textbf{system design} (that is, architecture modeling), MetaGPT employs its \textit{Architect} agent to derive and visualize the system architecture. The \textit{Architect} agent operates on the product requirements document produced by the \textit{Product Manager} agent, and the \textit{Architect} agent can generate two mermaid files that record the UML syntax of a class diagram and a sequence diagram, respectively, and a brief \texttt{Implementation Approach} description and \texttt{Anything UNCLEAR} declaration. Figure~\ref{fig:CDcomparison} and Figure~\ref{fig:SDcomparison} present a comparison of the class and sequence diagrams produced by MAAD and MetaGPT, respectively. To enable visualization of the UML diagrams produced by both systems, we employed PlantUML\footnote{https://plantuml.com/} to convert the textual UML specifications into graphical representations. In particular, this visualization capability is integrated directly into the MAAD framework, allowing it to automatically generate visual UML diagrams from its generated artifacts. %In Figure~\ref{fig:SDmetagpt}, the abbreviations denote the following components: SF for SpaceFractions, G for Game, A for Admin, Q for Questions, FO for FractionOperations, F for Feedback, and QM for QuestionManager. 

In terms of \textbf{documentation}, MetaGPT produces one JSON file to record technical solutions, including the following fields: 

\begin{itemize}
    \item \textit{Required Python Packages} include the necessary Python packages.
    \item  \textit{Required Other language third-party packages} refer to the selection of technology stack, which belongs to the technology dependency decision.
    \item \textit{Logic Analysis} defines the division of responsibilities of modules or files (e.g., \textsc{game.js} handles game logic, \textsc{admin.js} manages backend functions), reflecting the system architecture design.
    \item \textit{Task List} presents the code files to be implemented, which belong to the development task split.
    \item \textit{Full API Spec} describes API design specification.
    \item \textit{Shared Knowledge} describes the general design principles of the system (e.g., front-end framework selection, class or function sharing mechanism), which belongs to the architecture constraint description.
    \item \textit{Anything UNCLEAR} identifies issues that need to be clarified (such as browser compatibility, administrator interface design), which belong to requirements defect tracking and provide input for subsequent iterations.
\end{itemize}

Ideally, all the fields of the MetaGPT JSON files should contain content, but in fact, the experimental results show that certain fields are null values, such as \texttt{Required Python packages} and \texttt{Full API spec}. Compared to MetaGPT, MAAD generates one architectural document that records detailed architecture design solutions (see Section~\ref{sec:DesignerAgent}). Furthermore, MetaGPT lacks the ability to comprehensively evaluate the quality of its generated architecture design. Consequently, we are unable to assess the differences between MAAD and MetaGPT that use architectural evaluation approaches executed by LLMs. To address this, we conducted interviews with industry practitioners. In general, we summarized the results of the comparison between MAAD and MetaGPT with respect to the automated architecture design in Table~\ref{tab:MAS_Comparison}.

\begin{table}[ht]
    \centering
    \footnotesize
    \caption{Comparison between MAAD and MetaGPT}\label{tab:MAS_Comparison}
    \begin{tabularx}{\textwidth}{|l|>{\RaggedRight\arraybackslash}p{0.2\textwidth}|>{\RaggedRight\arraybackslash}p{0.2\textwidth}|>{\RaggedRight\arraybackslash}p{0.2\textwidth}|>{\RaggedRight\arraybackslash}X|}
        \hline
        \textbf{MAS} & \textbf{ASR Extraction} & \textbf{Architecture Model Integrity} & \textbf{Documentation granularity} & \textbf{Architecture Evaluation} \\\hline
        MAAD & 
        \faThumbsOUp \quad Extract ASRs and classify requirements based on the original requirements. & 
        \faThumbsOUp \quad Generate ``4+1'' architectural view models. & 
        \faThumbsOUp \quad Generate detailed architecture documentation. & 
        \faThumbsOUp \quad Generate ATAM architecture evaluation report and mismatch report. \\\hline
        MetaGPT & 
        \faThumbsOUp \quad Generate a software requirements specification. \newline 
        \faThumbsODown \quad Contain unreal requirements generated by LLMs. & 
        \faThumbsODown \quad Generate only class diagrams and sequence diagrams. & 
        \faThumbsODown \quad Generate a simple technical solution. & 
        \faThumbsODown \quad No evaluation mechanism. \\\hline
    \end{tabularx}
\end{table}

\subsubsection{Interviews with Architects}\label{sec:RQ1_p3}

To further evaluate the performance of MAAD, we interviewed three practitioners to get their viewpoints. Three main themes emerged: reasons for positive ratings (\faThumbsOUp), reasons for negative ratings (\faThumbsODown), and suggestions for improving generations (\faLightbulbO).

\faThumbsOUp \quad \textbf{Correctness}. Participants agreed that MAAD's generated architectural view models align with established design principles, even though they are less complex than those found in industrial systems. Three participants conveyed the same viewpoint, ``\textit{the generated architecture view models confirm the principles of architecture design, although the models are not as complex as industrial systems}'' (P1, P2, P3).

\faThumbsOUp \quad \textbf{Practicability}. All interviewees reported that MAAD is highly useful for assisting architects, particularly through its detailed mismatch reports. P1 emphasized that: ``\textit{MAAD is undoubtedly useful to assist architects, especially mismatch reports. Its strength lies in the utilization of extensive external knowledge: the human brain has limited memory, whereas an LLM can schedule and recall knowledge much more comprehensively and rapidly}''. Moreover, the knowledge-driven feature of MAAD is highly rated by two participants, ``\textit{External knowledge libraries can better support customized domain-specific software design}'' (P1, P3).

\faQuestionCircleO \quad \textbf{Trustworthiness}. Despite acknowledging MAAD's practical benefits and potential, participants remained cautious about relying solely on LLM-generated artifacts. One participant remarked: ``\textit{Trustworthiness and explainability issues persist — not because of MAAD per se, but as a general challenge for all LLM outputs, especially in safety-critical domains}'' (P1). 

\faThumbsOUp \quad \textbf{Limitations in Correctness}. An interviewee flagged specific inaccuracies in the MAAD's performance. For example, ``\textit{MAAD extracts certain non-functional requirements and links them to quality attributes (e.g., maintainability), but these associations sometimes lack clear justification and may be incorrect}'' (P1).

\faThumbsODown \quad \textbf{Level of Detail}. Several participants noted that the generated models lack sufficient granularity. One participant commented: ``\textit{The generated architectural view models omit detailed UML descriptions, and the relationships between entities are not always represented}'' (P1). Another participant similarly noted: ``\textit{Compared with industrial applications, the complexity and granularity of requirements analysis and architecture documentation do not meet industrial standards}'' (P2).

\faLightbulbO \quad \textbf{Suggestions}. %Regarding output formats, P1 recommended using Markdown instead of plain text files (\texttt{.txt}) for better readability. 
P2 and P3 proposed that future multi-agent systems could assign each agent a specialized LLM, for instance, using a code-focused LLM for programmer agents, and equipping fine-tuned and domain-specific LLMs for other agent roles (P2, P3). Regarding reuse, P3 mentioned that ``\textit{agent design should consider adding a memory mechanism and reusing the architectural design based on previous business scenarios}'' (P3).

\faLightbulbO \quad \textbf{Challenges}. Assessing MAAD's \textit{Evaluator} agent remains challenging due to the subjective nature of design; as one participant observed, ``\textit{Design solutions that satisfy most requirements are good designs, but human evaluators can exhibit bias; architects often favor their own solutions over those generated by LLMs}''. This participant also highlighted the enduring issue of tacit knowledge: ``\textit{During software architecture design, much tacit knowledge is hard to capture and cannot yet be leveraged, which remains a major obstacle in the automated design driven by knowledge}'' (P1).

\begin{tcolorbox}
\textbf{RQ1 Summary.} \textit{The MAAD framework produces various artifacts that clearly demonstrate its effectiveness and advantages in system architecture design. Moreover, our comparative evaluation shows that MAAD delivers a more refined and nuanced architectural solution than the baseline multi-agent system development framework, MetaGPT.}
\end{tcolorbox}

\subsection{Results of RQ2}\label{sec:RQ2_Results}

In the MAAD framework, both the \textit{Modeler} and \textit{Designer} agents leverage reference knowledge stored in a vector database to generate artifacts. We selected the third (2012) and fourth (2021) editions of the book \textit{Software Architecture in Practice}~\cite{Bass2012SAP, Bass2021SAP} as authoritative external knowledge sources and embedded them into a vector database via Retrieval-Augmented Generation (RAG)~\cite{gao2024retrieval}. To evaluate the impact of external knowledge on the generation of architecture design by MAAD, we conducted a comparative analysis of the artifacts produced by MAAD using two distinct prompt templates: one incorporating reference knowledge and the other devoid of such knowledge. Here, we then compared (1) the \textbf{structural comparison} of the system design (via architectural views) produced by the \textit{Modeler} agent and (2) the \textbf{mismatch} rates reported by the \textit{Evaluator} agent.

\begin{figure}
    \centering \includegraphics[width=1\linewidth]{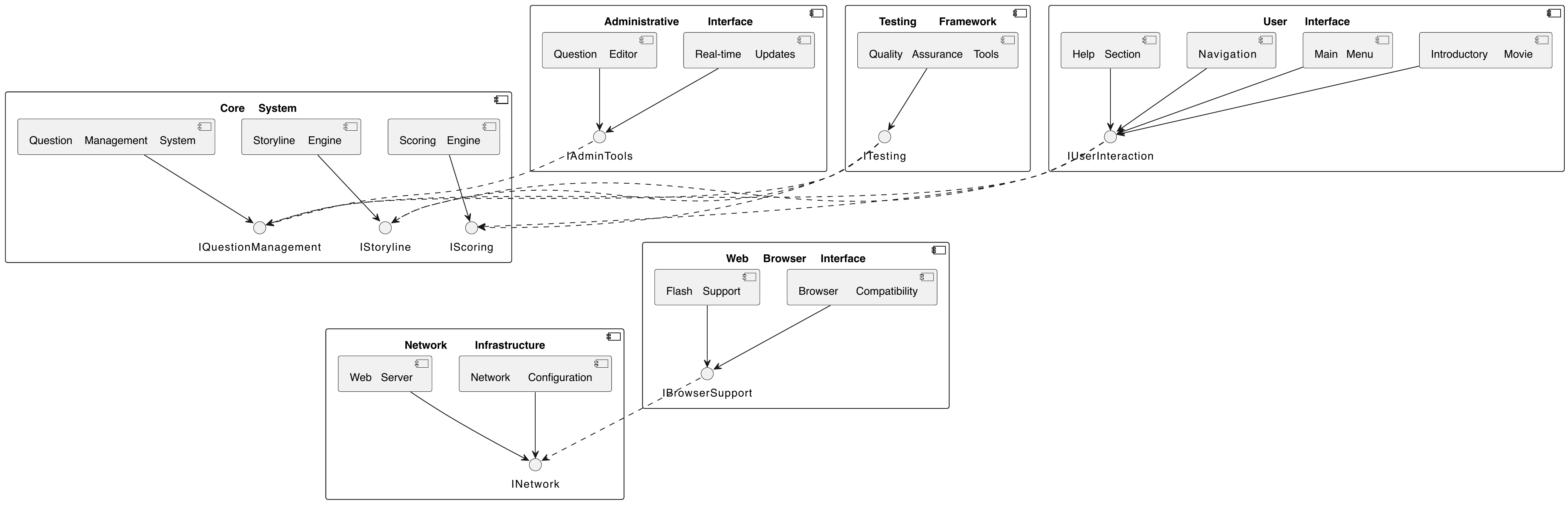}
    \caption{The component diagram of SFS with reference knowledge}
    \label{fig:component_view_withRAG}
\end{figure}

\begin{figure}
    \centering\includegraphics[width=1\linewidth]{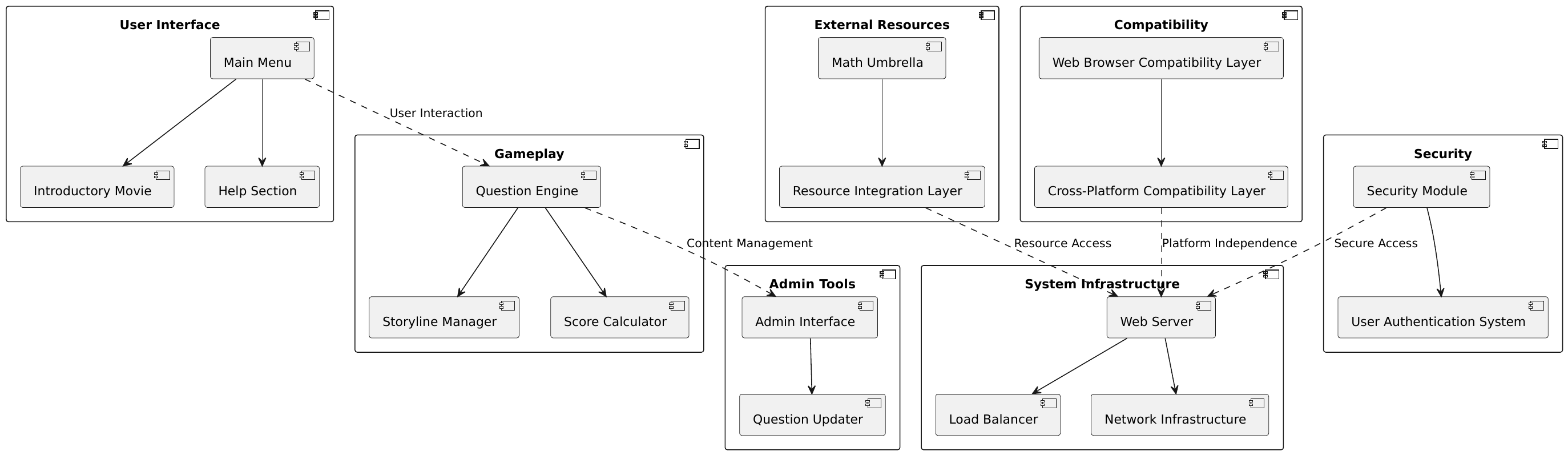}
    \caption{The component diagram of SFS without reference knowledge}
    \label{fig:component_view_withoutRAG}
\end{figure}

\textbf{Structural Comparison}. To answer RQ2, we still use the Space Fraction System (SFS) requirements as a running example. Figure~\ref{fig:component_view_withRAG} and Figure~\ref{fig:component_view_withoutRAG} respectively depict the component diagrams generated with and without external knowledge. We focus on the UML Component Diagram, since it clearly illustrates component decomposition and interface‐based abstraction at the design level. 
% with and without reference knowledge的区别
Figure~\ref{fig:component_view_withRAG} shows a modular, interface‐driven design: for instance, the \textsc{User Interface} component interacts with multiple front-end modules (e.g., \textsc{Main Menu}, \textsc{Help Section}, \textsc{Navigation}) via the \textsc{IUserInteraction} interface, thereby decoupling presentation logic from its implementations and enhancing maintainability. Within the core system, components such as \textsc{Scoring Engine}, \textsc{Storyline Engine}, and \textsc{Question Management System} communicate through \textsc{IScoring}, \textsc{IStoryline}, and \textsc{IQuestionManagement} interfaces, supporting encapsulation and facilitating future substitution. Explicit dependencies (e.g., \textsc{IUserInteraction} relies on \textsc{IScoring}) illustrate well‐defined collaboration paths and adhere to contract‐driven architectural principles. 
In contrast, Figure~\ref{fig:component_view_withoutRAG} illustrates a different component-level interaction diagram, emphasizing functional relationships and the execution flow among system modules. Despite this view clarifying execution sequences among system components, it lacks the abstraction and separation of concerns that interface‐oriented modeling provides. Consequently, Figure~\ref{fig:component_view_withoutRAG} aligns more closely with a Logical or Runtime View than a proper component view.

%Mismatch Rate的区别
\textbf{Mismatch Report}. In the mismatch report of SFS, the \textit{Evaluator} agent identifies six mismatches under both conditions (i.e., with or without external knowledge), yielding an identical mismatch rate of 0.188 in each case by Equation~\ref{eq:Mismatch Rate}. Through the infusion of external knowledge, the identified mismatches include reliance on outdated technologies (e.g., Adobe Flash), omission of essential functions (e.g., real‐time score calculation, dynamic storyline adaptation), and inadequate support for key non‐functional requirements (e.g., security and reliability), which is related to the quality of external knowledge. In contrast, when external knowledge is not provided, the mismatches are primarily associated with functional-detail discrepancies (e.g., inconsistent input‐method specifications), missing architectural support for dynamic content updates, and misalignment between non‐functional requirements and design elements (e.g., vague security requirements absent corresponding modules).

\begin{equation}
\label{eq:Mismatch Rate}
\text{Mismatch Rate} = \frac{\text{Number of Mismatches}}{\text{Total Number of Requirements}}
\end{equation}

Overall, integrating external knowledge into MAAD's vector database produced component diagrams that more closely align with established architectural principles, especially in component decomposition, separation of concerns, and interface design.

\begin{tcolorbox}
\textbf{RQ2 Summary.} \textit{Incorporating external knowledge into MAAD yields more modular and interface‐driven architecture design, aligning with architectural best practices. To some extent, external knowledge can help to improve the quality of architecture design.}
\end{tcolorbox}

% How do different LLMs affect the quality of architecture design in the MAAD framework?
\subsection{Results of RQ3}\label{sec:RQ3_Results}

\begin{figure}[htb]
    \centering \includegraphics[width=1\linewidth]{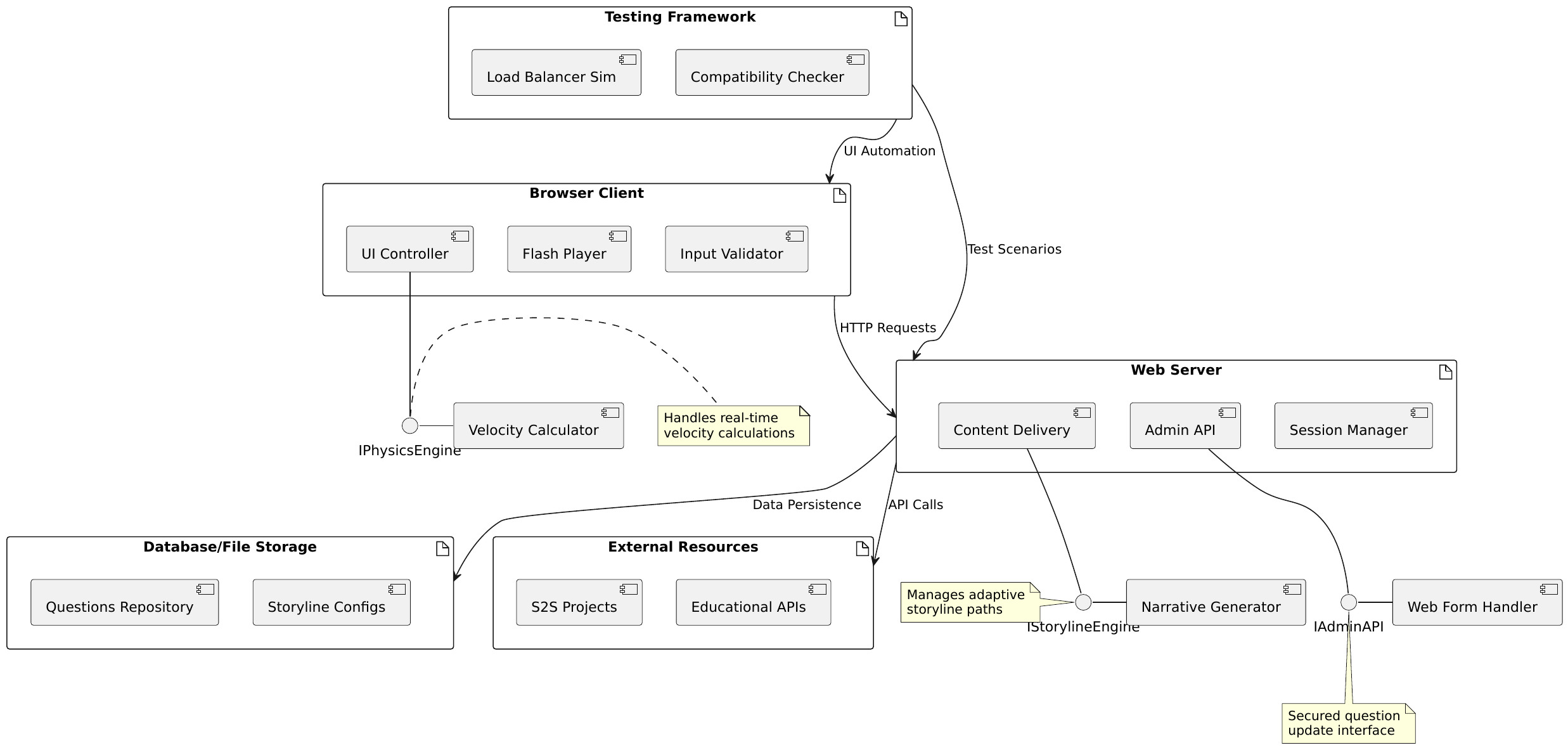}
    \caption{The component diagram of SFS generated by DeepSeek-R1}
    \label{fig:develop_view_deepseek}
\end{figure}

\begin{figure}[htb]
    \centering\includegraphics[width=0.58\linewidth]{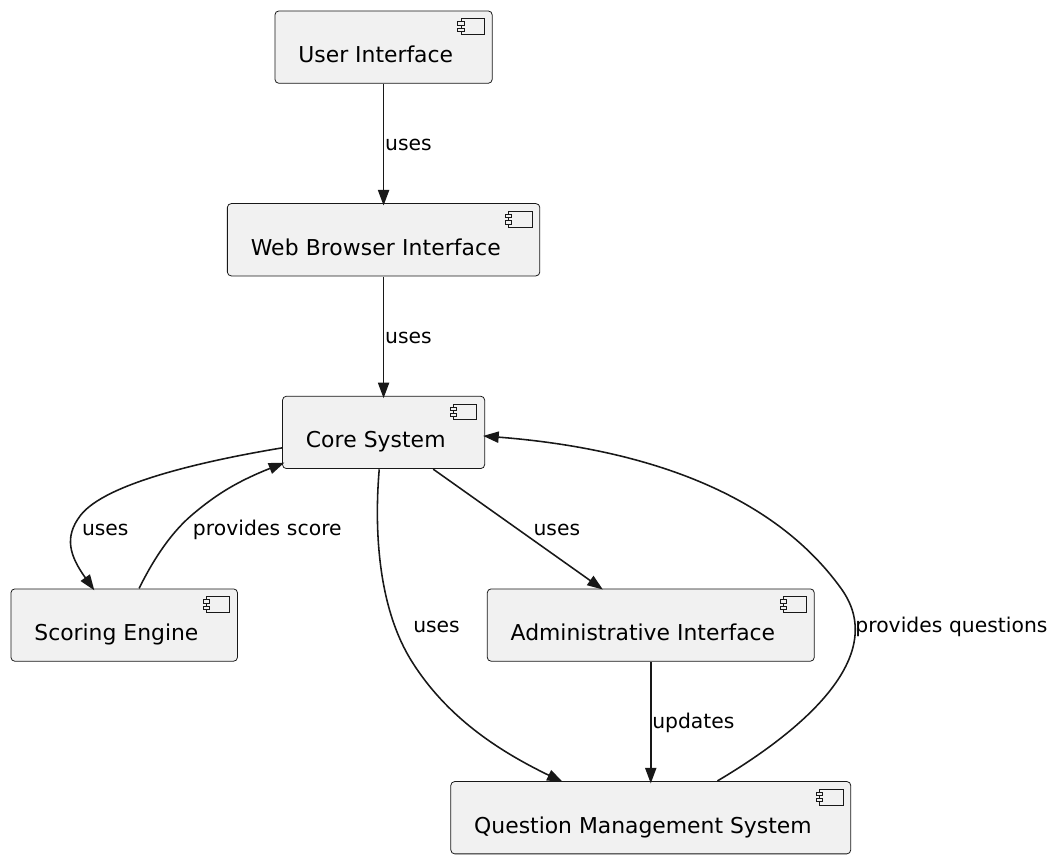}
    \caption{The component diagram of SFS generated by Llama 3.3}
    \label{fig:develop_view_llama3.3}
\end{figure}

To answer RQ3, we extended our evaluation beyond GPT-4o by generating SFS architecture designs with two additional LLMs, i.e., DeepSeek-R1 and Llama 3.3. Likewise, to conduct \textbf{structural and mismatches comparison}, we equipped MAAD with DeepSeek-R1 (671B) and Llama 3.3 (70B) as foundational LLMs to design the architecture using the same SRS input. Figure~\ref{fig:component_view_withRAG}, Figure~\ref{fig:develop_view_deepseek}, and Figure~\ref{fig:develop_view_llama3.3} depict the component diagrams of SFS by GPT-40, DeepSeek-R1 and Llama 3.3, respectively. 

\textbf{Structural Comparison}. The diagram generated by DeepSeek-R1 emphasizes concrete deployment and runtime aspects: it models explicit artifacts such as client, server, and database instances, and traces interaction paths (e.g., HTTP requests and API calls). Within each unit, it further decomposes functionality (e.g., UI Controller and Narrative Generator), which offers richer visibility into implementation-level responsibilities. In contrast, GPT-4o's representation centers on logical component dependencies and abstracts away most deployment details and internal module structure. By comparison, Llama 3.3 produces a streamlined, high-level component diagram, which highlights major subsystems (user interface, core system, administrative interface) and their data flows, but omits both fine-grained internal operations and deployment specifics. Moreover, {GPT-4o's} diagram tends to capture more detailed interface contracts and internal structures, and Llama~3.3's output favors conceptual clarity over implementation precision, making it more suitable for early-stage architectural planning rather than detailed design analysis.

In terms of the component diagrams generated by the three LLMs, all three LLMs yielded functionally plausible architectures within the MAAD framework. Although variations exist across architectures, such differences are reasonable, as software requirements can be satisfied by multiple valid architectural solutions.

\textbf{Mismatch Report}. As for the mismatch reports generated by the \textit{Evaluator} agent, six requirement-architecture mismatches (mismatch rate of 0.188) were produced by MAAD with GPT-4o and Llama 3.3 as the base LLMs, including gaps in cross-browser compatibility, maintainability, real-time update strategies, security provisions, administrative interface design, and scalability planning. 
% Mismatch report的内容可以不用写
%The analysis of the system requirements and architecture design reveals several key mismatches: (1) The architecture design lacks a clear plan for ensuring cross-browser compatibility, potentially leading to accessibility and compatibility issues; (2) The design does not address maintainability, risking challenges in updating and modifying the system; (3) The approach for implementing real-time content updates is insufficiently detailed, which could cause disruptions and downtime; (4) Security considerations, particularly regarding user data and authentication, are not adequately covered; (5) The design fails to specify how to create a user-friendly administrative interface for managing content; and (6) There is insufficient detail on scalability and fault-tolerance, which may hinder the system's ability to handle increased traffic or failures. These mismatches suggest a gap between the system's requirements and its architectural implementation, impacting the system's overall reliability, security, usability, and performance.
The architecture design produced by DeepSeek-R1 exhibited ten mismatches, yielding a mismatch rate of 0.313. These mismatches can fall into five categories, including unsupported security mechanisms, outdated technologies, ambiguous adaptivity definitions, missing multi-admin consistency checks, and misaligned enhancements. The results indicate a weaker alignment with the SRS.
%: (1) security mechanisms such as HTTPS and SHA-256 are implemented in the architecture but lack corresponding support in the SRS; (2) outdated technologies (e.g., Flash) or conflicting deployment assumptions are used; (3) the definitions of system adaptivity and narrative mechanisms are vague or incomplete; (4) the architecture lacks consistency safeguards for multi-admin editing workflows; and (5) several enhanced features—such as scalability and security hardening—are present in the architecture but are not clearly mapped to specific requirements.

Overall, for the SFS requirements case, the mismatches in the architecture design generated by DeepSeek-R1 are higher than those in GPT-4o and Llama 3.3. This indicates that the selection of basic LLMs for MASs can significantly impact requirements coverage and architectural consistency.

\begin{tcolorbox}
\textbf{RQ3 Summary.} \textit{In the case of SFS, although GPT-4o, DeepSeek-R1, and Llama 3.3 all produce functionally plausible SFS architectures, DeepSeek-R1's design incurs a substantially higher mismatch rate. The choice of basic LLMs within MAAD has a marked impact on both requirements coverage and architectural consistency.}
\end{tcolorbox}

\section{Discussions}\label{sec:Discussions_and_Implications}

In this section, we explain the experimental results and outline the implications for subsequent practices and research.

\subsection{Interpretations}
\subsubsection{Interpretations on RQ1 Results}
% RQ1 部分：谈论MAAD跟MetaGPT对比的结构，优势和劣势
% Interview结果的进一步讨论。
% 例如，现有的MAS在纯软件系统相关的项目时，可能更有优势，一旦有些软件系统需要调用硬件设备（需要跟硬件交互），这时MAS的局限性就凸显出来了？

According to the results of RQ1, we found that MAAD is capable of generating architecture designs that are both plausibly complete and highly relevant, as demonstrated through the case study of the SFS requirements. \textbf{Compared to MetaGPT, MAAD exhibits superior performance in analyzing and categorizing user requirements, as well as in producing more fine-grained and comprehensive architectural solutions}. These include detailed documentation and support for the ``4+1'' architecture view models, which enhance \textit{traceability} (i.e., the ability to trace architectural elements back to original requirements) and \textit{architectural consistency} (i.e., the internal alignment and integration among different architecture views and design artifacts).

MetaGPT is designed to automate the entire software development lifecycle, transforming user requirements directly into executable code. In contrast, MAAD is specifically intended for generating architecture designs, with a distinct emphasis on high-level structural design. Their methodologies and design emphases diverge considerably. MAAD emphasizes in-depth requirements analysis, architectural integrity, and systematic evaluation of architectural quality. MetaGPT, on the other hand, operates as an end-to-end software development MAS along with orchestrated Standard Operating Procedures (SOPs), focusing on delivering complete implementations. Due to its specialization in architectural design, MAAD prioritizes robustness of the architectural design process and long-term system maintainability, making it more suitable for projects where system reliability and architectural completeness are critical.

\subsubsection{Interpretations on RQ2 Results}

The results of RQ2 show that \textbf{the infusion of external knowledge enables MAAD to produce more abstract, interface‐driven component diagrams that align more closely with established architectural best practices}. In contrast, when external knowledge is absent, the generated diagrams tend to resemble lower-level runtime views. These qualitative improvements highlight the importance of domain grounding in architectural design generation. However, the \textit{Evaluator} agent's mismatch rate remains unchanged at 0.188, which implies that the mismatch rate might be influenced by multiple factors, as architectural design involves balancing various requirements and quality attributes. One possible reason could be that the fundamental LLMs used in MAAD already contain architectural knowledge, as previously demonstrated by Soliman \textit{et al}. \cite{Soliman2025LLMAK}. This implies that the bottleneck may lie in domain-specific knowledge gaps rather than in a lack of general architectural competence.

Owing to its knowledge-driven feature, MAAD can integrate external knowledge sources, including authoritative literature and proprietary domain-specific knowledge databases. This extensibility supports the integration of specialized domain knowledge, enabling more precise tailoring of architecture generation to the needs of specific application domains. By grounding its reasoning in curated knowledge, we believe that MAAD could help to increase trustworthiness and consistently deliver higher-quality designs with minimal human oversight.

% Future work should focus on expanding the breadth and depth of domain-relevant knowledge integration. This includes developing mechanisms for dynamic knowledge ingestion and validation to further improve the contextual relevance and precision of generated architectural artifacts.

\subsubsection{Interpretations on RQ3 Results}

The results of RQ3 reflect that LLM-generated architectures are not merely ``correct'' or ``incorrect'' but reflect inherent priorities and differences during the training of LLMs. Compared to GPT-4o and Llama 3.3, the relatively higher mismatch rate observed in DeepSeek-R1's outputs suggests that LLMs vary in their capacity to produce abstract system designs. The possible reason could stem from differences in training data and optimization goals. 

Moreover, the results also underscore that \textbf{an LLM's architectural design capability is closely tied to how well its intrinsic modeling capabilities align with the comprehension of requirements}. This finding highlights the importance of LLM selection not solely based on conventional performance metrics, but also on how the model's reasoning ability complements the specific phases and goals of the software architecting process.

Furthermore, improving design quality may require model calibration techniques such as reinforcement learning from architectural evaluation feedback or the integration of architectural knowledge (e.g., architecture patterns and principles
). Furthermore, the results emphasize the need for task-aware LLM selection within multi-agent systems: different LLMs may be more effective when matched to specific architectural activities (e.g., architectural understanding, evaluation, and implementation), thereby mitigating design gaps and enhancing overall design quality.

\subsection{Implications}
% \textbf{Certain development activities cannot be automated.} 尤其是跟需求相关的活动
% Kang and Shaw \cite{Kang2024AIwndSE} assert that while Generative AI could be a valuable support tool for tasks like requirements elicitation, design, and reliability engineering, it is improbable that it will fully automate or replace them. These tasks are heavily influenced by context and require flexibility, judgment, common sense, and a significant amount of tacit knowledge.
% Architects should be on standby to intervene when the LLM-based agents' outputs are unsatisfactory. 
% There is still much to be studied in the multi-agent coordination algorithms, although the latest Model Context Protocol and Agent2Agent protocol have attracted more attention.
% tacit knowledge很难捕捉和利用

Based on the findings, we outline several key implications for the future of automated software architecture design.

\textbf{Multi-Agent Systems Provide Superior Performance for Software Architecting Tasks.} Our analysis in RQ1 shows that MAAD's specialized four-agent framework substantially outperforms MetaGPT's single-agent setup in generating high-quality architectural designs. While MetaGPT generated only basic class and sequence diagrams, MAAD produced comprehensive artifacts spanning complete ``4+1'' architectural views, detailed documentation, and systematic evaluation. The results suggest that for software architecting tasks demanding collaborative reasoning and varied expertise, researchers and tool developers should give priority to MASs with specialized roles rather than single, general-purpose agents.

\textbf{Knowledge Integration Improves Quality but Requires Domain Expertise.} In RQ2, while external knowledge integration produced more modular, interface-driven component diagrams that better align with architectural best practices, the unchanged mismatch rate (0.188) reveals a critical limitation: while generic architectural knowledge might provide qualitative improvements, it fails to deliver domain-specific design without domain context. This finding indicates that organizations deploying knowledge-driven automation must invest in building domain-specific knowledge bases tailored to their application contexts rather than relying exclusively on the general software engineering literature to achieve measurable performance gains. 

\textbf{LLM Selection in MAAD Critically Impacts Architecture Design.} Building on the multi-agent foundation, our RQ3 results reveal that the choice of underlying LLM significantly affects output quality, with GPT-4o and Llama 3.3 achieving 0.188 mismatch rates compared to DeepSeek-R1's 0.313 rate, reflecting a 66\% drop in requirements alignment. Notably, DeepSeek-R1, despite being considered a stronger model overall, performed worse than Llama 3.3 on architecture design tasks. This counterintuitive finding indicates that organizations implementing automated software engineering systems should conduct task-specific LLM evaluations rather than relying on general model rankings or capabilities, as model performance can vary significantly across different types of reasoning and generation tasks.

\textbf{Human Validation Remains Essential for Architecture Decisions.} Expert interviews consistently positioned MAAD as ``\textit{undoubtedly useful for assisting architects, especially mismatch reports}'' while emphasizing that trustworthiness and explainability issues persist. Practitioners appreciated MAAD's ability to recall knowledge comprehensively and perform systematic analysis, yet emphasized the continued need for human validation, particularly noting that ``\textit{architects often favor their own solutions over those generated by LLMs}''. This feedback demonstrates that organizations should position automated architecture design tools as assistant technologies requiring human validation rather than autonomous systems replacing architects, especially in safety-critical domains where architectural decisions have significant consequences.

\section{Threats on Validity}\label{sec:Threats}

One of the threats is the potential hallucinations of LLMs~\cite{Zhang2025Hallucination}. To mitigate the threat in architecture design, we employed a multi-agent system where agents cross-verify each other's generations, reducing the likelihood of spurious or incorrect designs. Additionally, we extracted extensive design knowledge from trustworthy sources, including well-established architectural patterns and best practices, to ground the generated designs in reliable information. To further enhance quality assurance, we incorporated an LLM-based evaluator that assesses the reasonableness of the proposed designs, ensuring that the generated artifacts align with expected architectural principles and domain requirements.

Another threat concerns the domain specificity of our knowledge database. To address this limitation, we integrated authoritative literature and employed vector-based retrieval to identify relevant knowledge segments. However, as evidenced in our RQ2 results, the current knowledge base lacks domain-specific content for educational software (the domain that the case project SFS belongs to), which may constrain MAAD's ability to generate highly specialized and context-aware designs. To mitigate this limitation, future research can incorporate domain-specific knowledge into MAAD to enhance its contextual understanding and adaptability.

The third threat pertains to the evaluation of architecture quality, which remains inherently subjective. To address this, we employed multiple complementary evaluation approaches: systematic mismatch analysis, ATAM-based evaluation reports, and validation through semi-structured interviews with experienced architects. While these methods provide valuable insights, we recognize that architecture evaluation cannot be entirely objective. Consequently, the current assessment may not fully capture the quality or effectiveness of the generated architecture designs.
\section{Conclusions and Future Work}\label{sec:Conclusion}

In this study, we proposed an automated software architecture design framework, Multi-Agent Architecture Design (MAAD), a knowledge-driven Multi-Agent System (MAS) specifically developed to autonomously generate architectural designs along with comprehensive architecture evaluation reports assessing architectural quality. MAAD comprises four specialized agents (i.e., \textit{Analyst}, \textit{Modeler}, \textit{Designer}, and \textit{Evaluator} agent) to collaboratively generate system architecture design based on given software requirements specifications.

To evaluate the utility and correctness of the MAAD framework, we conducted a case study showing the detailed generated artifacts of each agent. We compared MAAD with MetaGPT, a state-of-the-art software development MAS, highlighting MAAD's ability to produce not only more comprehensive and complete architectural artifacts but also architecture evaluation and mismatch reports. Additionally, feedback was collected from industry architects based on MAAD's performance for 11 real-world SRSs, and the responses were largely positive, reinforcing MAAD's practical applicability. Our findings indicate that incorporating external domain knowledge can enhance the quality of the generated architectural artifacts. Furthermore, we evaluated the performance of MAAD under three foundational LLMs, namely GPT-4o, DeepSeek-R1, and Llama~3.3. The results reveal that GPT-4o significantly outperforms the others regarding the quality of generated artifacts, underscoring the influence of the underlying LLM on the overall performance of MAS.

Despite the promising outcomes, we recognize that architecture design remains a complex and iterative task. While MAAD represents a meaningful step toward the automation of architecture design, its current capabilities might be limited when applied to highly complex industrial systems. In the next steps, we aim to improve MAAD by integrating memory mechanisms and orchestrating a more granular set of agents to better address intricate and large-scale software requirements.

\section*{Data Availability}\label{sec:DataAvailability}

We have made the prompts for agent setting, interview protocol, scripts of MAAD, and experimental outputs available in our replication package~\cite{onlinepackage_TOSEM}.

% \begin{acks}
% This research is supported by the National Natural Science Foundation of China (NSFC) with Grant No. 62402348 and 62172311; National Research Foundation, Prime Minister's Office, Singapore under the Campus for Research Excellence and Technological Enterprise (CREATE) programme; the National Research Foundation, Singapore, and DSO National Laboratories under the AI Singapore Programme (AISG Award No: AISG2-GC-2023-008). Besides, the authors would like to thank all participants in this study.
% \end{acks}

% \clearpage
\bibliographystyle{IEEEtran}
\bibliography{TOSEM_ref}

\end{document}